\documentclass[journal]{IEEEtran}
 \usepackage[cmex10]{amsmath}
 \usepackage{eqparbox}
 \usepackage[usenames, dvipsnames]{color}
 \usepackage{mathtools,amssymb,bm,mathabx}
 \usepackage{amstext}
 \usepackage{amssymb}
 \usepackage{graphicx}
 \usepackage{color}
 \usepackage{booktabs}
 \usepackage{longtable}
 \usepackage{multicol}
 \usepackage{amsfonts}
 \usepackage{dsfont}
 \usepackage{array}
 \usepackage{algorithmicx}
 \usepackage[ruled]{algorithm}
 \usepackage{algpseudocode}
 \usepackage{algpascal}
 \usepackage{algc}
 \usepackage{cases}
 \usepackage{pgfplots}
 \usepackage{accents}
 \usepackage{caption}
 \usepackage[footnotesize]{subfigure}
 \usepackage{amsthm,xparse}
 \usepackage{float}
 \usepackage{cite}
 \usepackage [autostyle, english = american]{csquotes}
 \usepackage{hyperref}
 \theoremstyle{remark}

 \newcommand\ASTART{\bigskip\noindent\begin{minipage}[b]{0.5\linewidth}}
 	
 	\newcommand\AENDSKIP{\end{minipage}\bigskip}
 \newcommand\AEND{\end{minipage}}
\ifCLASSOPTIONcaptionsoff
\usepackage[nomarkers]{endfloat}
\let\MYoriglatexcaption\caption
\renewcommand{\caption}[2][\relax]{\MYoriglatexcaption[#2]{#2}}
\fi
\theoremstyle{plain}
\newtheorem{thm}{\textbf{Theorem}}
\newtheorem{lem}{\textbf{Lemma}}
\newtheorem{prop}{\textbf{Proposition}}
\newtheorem{corl}{\textbf{Corollary}}

\theoremstyle{definition}
\newtheorem{defn}{\textbf{Definition}}

\theoremstyle{remark}
\newtheorem*{rem}{\bf Remark}

\newcommand*{\rom}[1]{\expandafter\@slowromancap\romannumeral #1@}
\def\change{black}
\newcommand{\RN}[1]{%
\textup{\uppercase\expandafter{\romannumeral#1}}%
}

\usepackage{standalone}
\graphicspath{{./Figures/}} 
\begin{document}
%
\title{Exploiting Prior Information in Block-sparse Signals}
\author{Sajad~Daei, Farzan~Haddadi, Arash~Amini}%

\maketitle

\begin{abstract}
We study the problem of recovering a block-sparse signal from under-sampled observations. The non-zero values of such signals appear in few blocks, and their recovery is often accomplished using an $\ell_{1,2}$ optimization problem. In applications such as DNA micro-arrays, some extra information about the distribution of non-zero blocks is available; i.e., the number of non-zero blocks in certain subsets of the blocks is known. A typical way to consider the extra information in recovery procedures is to solve a weighted $\ell_{1,2}$ problem. In this paper, we consider a block-sparse model which is accompanied with a partitioning of the blocks; besides the overall block-sparsity level of the signal, we assume to know the block-sparsity of each subset in the partition. Our goal in this work is to minimize the number of required linear measurements for perfect recovery of the signal by tuning the weights of a weighted $\ell_{1,2}$ problem. For this goal, we apply tools from conic integral geometry and derive closed-form expressions for the optimal weights. We show through precise analysis and simulations that the weighted $\ell_{1,2}$ problem with optimal weights significantly outperforms the regular $\ell_{1,2}$ problem. We further show that the optimal weights are robust against the inaccuracies of prior information.
\end{abstract}


%
\IEEEpeerreviewmaketitle

\section{Introduction}
\begin{figure}[t]
	\hspace{-.3cm}
	\includegraphics[scale=.25]{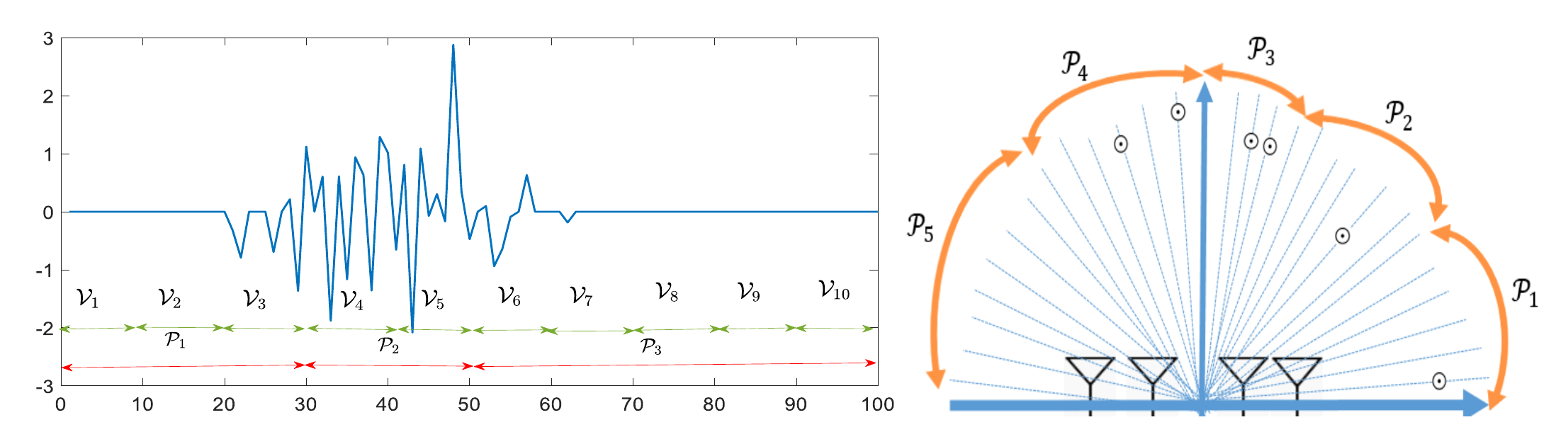}
	\caption{Left image: Illustration of a non-uniform block-sparse signal $\bm{x}\in\mathbb{R}^n$ with parameters $n=100$, $q=10$, $k=10$, $L=3$, $\alpha_1=\tfrac{1}{3}$, $\alpha_2=1$ and $\alpha_3=\tfrac{2}{5}$. Right image: Schematic diagram of DOA estimation of far-field sources. The angular half-space is divided into $q=30$ angular clusters $\{\mathcal{V}_b\}_{b=1}^q$ with equal length. $L=5$ block support estimators with accuracies $\alpha_1=\tfrac{1}{6}$, $\alpha_2=\tfrac{1}{6}$, $\alpha_3=\tfrac{1}{2}$, $\alpha_4=\tfrac{1}{3}$ and $\alpha_5=0$ are considered.}
	\label{fig.model}
\end{figure}

 \IEEEPARstart{C}{ompressed} sensing (CS) aims at recovering a sparse signal $\bm{x}\in\mathbb{R}^n$ from a few random linear measurements
 \begin{align}\label{eq.linear_measure}
 \bm{y}=\bm{A}\bm{x}+\bm{e},
 \end{align}
 where $\bm{A}\in \mathbb{R}^{m\times n}$ is called the measurement matrix, $\bm{e}$ is the noise vector which is assumed to be i.i.d. Gaussian with variance $\sigma_{e}^2$, and $m$ is much smaller than $n$. In this work, the signal is assumed to be block-sparse which occurs in many applications such as DNA micro-arrays \cite{stojnic2008reconstruction}, direction of arrival (DOA) estimation \cite{hyder2010direction}, computational neuroscience \cite{computationalnero}, and multiple measurement vector (MMV) problem \cite{zhu2017performance}. Consider a block-sparse signal $\bm{x}\in\mathbb{R}^{n}$ which is a concatenation of $q$ blocks $\mathcal{V}_b, b=1,...,q$. 
 The block support is defined as the index set of non-zero blocks, which we denote by $\mathcal{B}$. Figure \ref{fig.model} shows an example of a block-sparse signal composed of $q=10$ blocks {\color{\change} of length $k=10$}. 
 The block support in this figure is the index set $\{3,\dots, 7\}$; i.e., the signal is $5$-block-sparse and contains zero blocks except for $\{\mathcal{V}_3,\dots, \mathcal{V}_7\}$. 
 Ideally, a block-sparse signal is reconstructed using the following optimization problem:
 \begin{align}\label{P02}
 \mathsf{P}_{0,2}^{\eta}: &\min_{\bm{z}\in\mathbb{R}^n}\|\bm{z}\|_{0,2}:=\sum_{b=1}^{q}\bm{1}_{\|\bm{z}_{\mathcal{V}_b}\|_2>0}\nonumber\\
 & \mathrm{s.t.}~~ \|\bm{y}-\bm{A}\bm{z}\|_2\le \eta,
 \end{align}
 where $\bm{1}_{\mathcal{E}}$ denotes the indicator function of the event $\mathcal{E}$, and $\eta$ is an upper-bound for $\|\bm{e}\|_2$. This problem is computationally intractable in polynomial time and in general, it is NP-hard. Following Donoho \cite{donoho2006most}, $\mathsf{P}_{0,2}^{\eta}$ can be relaxed as an $\ell_{1,2}$ minimization of the form
\begin{align}\label{P12}
\mathsf{P}_{1,2}^{\eta}: &\min_{\bm{z}\in\mathbb{R}^n}\|\bm{z}\|_{1,2}:=\sum_{b=1}^{q}\|\bm{z}_{\mathcal{V}_b}\|_2\nonumber\\
& \mathrm{s.t.}~~ \|\bm{y}-\bm{A}\bm{z}\|_2\le \eta.
\end{align}
It is known that $\mathsf{P}_{1,2}^{\eta}$ for $\eta=0$ (noiseless case) finds the original block-sparse signal with high probability if $\bm{A}$ comes from a probability law that distributes the null space  uniformly with respect to the Haar measure\cite{amelunxen2013living}. This includes Gaussian ensembles and partial Fourier matrices. We also assume the same type of random matrices in this paper.

The main challenge in recovering a block-sparse signal is to find the block support. Intuitively, if there exists an additional piece of information about the block-sparsity level of certain subsets of the blocks\footnote{Our block-sparse model is deterministic in this paper; the normalized block-sparsity of a subset of blocks (the number of nonzero blocks divided by the total number of blocks in the subset) is occasionally referred to as the non-zero probability or non-zero likelihood of the blocks in this subset.}, 
one can probably solve $\mathsf{P}_{1,2}^{0}$ and $\mathsf{P}_{1,2}^{\eta}$ with fewer measurements or smaller reconstruction error, respectively (see Subsection \ref{sec.app} for more explanations). In some scenarios, such extra information is available. For example, in DNA microarray one might know 
the occurrence frequency of
 specific Genes in certain sets of arrays. In computational neuroscience, the behavior of neurons exhibit non-uniform clustered responses \cite{computationalnero}. In radar signal processing, an operator might know the range of speed with which an aircraft flies or the range of angular domain it forms in the radar \cite{mishra2015spectral}. 
 A typical way to consider such prior information is to apply a weighted $\ell_{1,2}$ minimization formulated as:
\begin{align}\label{P12w}
\mathsf{P}_{1,2,\bm{w}}^{\eta}: &\min_{\bm{z}\in\mathbb{R}^n}\|\bm{z}\|_{1,2,\bm{w}}:=\sum_{b=1}^{q}w_b\|\bm{z}_{\mathcal{V}_b}\|_2\nonumber\\
& \mathrm{s.t.}~ \|\bm{y}-\bm{A}\bm{z}\|_2\le \eta,
\end{align}
where $\bm{w}=[w_1,...,w_q]^T$. This work explicitly takes prior block information into account by optimally tuning the weights in $\mathsf{P}_{1,2,\bm{w}}^{\eta}$. Specifically, the main objective of this paper is to find optimal weights in $\mathsf{P}_{1,2,\bm{w}}^{\eta}$ in the sense that they minimize the reconstruction error for $\eta>0$, and the required number of measurements for $\eta=0$.

\subsection{Contributions}
Although block-sparse signals are widely used in many applications in the literature (see Subsection \ref{sec.related}), the inclusion of prior information about the block-support distribution in improving the performance of the signal recovery is not analyzed yet. In this work, we first consider the standard block-sparse model without prior information and derive tight and non-asymptotic bounds for the required sample complexity of $\mathsf{P}_{1,2}^{0}$. Interestingly, the asymptotic form of our bound matches the conjecture in \cite{donoho2013accurate}.
Then, we consider the weighted $\mathsf{P}_{1,2}^{\eta}$ problem  denoted by 
$\mathsf{P}_{1,2,\bm{w}}^{\eta}$ in presence of prior information for which we present a new method for obtaining the optimal weights. 
In particular, we assume to know a partitioning of $[q]:=\{1,\dots,q\}$ into $L$ disjoint subsets $\{\mathcal{P}_i\}_{i=1}^{L}$, about which we know the number of non-zero blocks in each index set $\mathcal{P}_i$. In simple words, we assume to know the block-sparsity level of each subset in a partitioning of the blocks. 
These partial block-sparsity levels are commonly expressed via $\alpha_i:=\tfrac{|\mathcal{P}_i\cap \mathcal{B}|}{|\mathcal{P}_i|}$ which are called the accuracy of $\mathcal{P}_i$s.
Obviously, if the partition consists of a single subset $[q]$, we are dealing with a conventional block-sparse model without prior information. Note that the measurement matrix combines all the blocks before generating the measurement vector. Hence, we do not have access to measurements that are solely determined by the blocks of a single partition; otherwise, the recovery problem could be simplified to multiple block-sparse recovery in lower dimensions without prior information.
We call signals generated based on this model as \textit{non-uniform} block-sparse signals. The left image of Figure \ref{fig.model} shows a sample signal with this model.
The strategy of finding optimal weights is to study the asymptotics of the upper and lower bounds of the statistical dimension of a certain convex cone (these bounds differ from each other in an asymptotically vanishing constant term). In this regard, we prove that our method of obtaining optimal weights is robust to slight inaccuracies of the prior knowledge. 
With optimal weights in $\mathsf{P}_{1,2,\bm{w}}^{0}$, we also find that the number of measurements required for exact recovery of a block-sparse vector equals to the sum of the required number of measurements in the simpler (smaller size) problems of recovering each subset via $\mathsf{P}_{1,2}^{0}$ separately.


\subsection{Related works and Key Differences}\label{sec.related}
There are extensive works in CS literature that consider reconstructing low-dimensional signals with prior information; for instance see \cite{khajehnejad2011analyzing,friedlander2012recovering,needell2017weighted,Baraniuk2010,xu2010compressive,bereyhi2018maximum}. Here, we discuss some of them.

The weighted $\ell_1$ minimization for sparse signal recovery is likely initiated by the work \cite{candes2008enhancing}. While this work reveals the advantages of weighting, it does not consider any prior information about the distribution of the support. In \cite{baraniuk2010model}, partial knowledge about the support is assumed as prior information. In this work, some of the standard sparse recovery techniques are modified to benefit from the available prior information; however, none of the methods rely on weighting.
%
A non-uniform sparse structure similar to the one that is considered in this paper, is proposed in \cite{khajehnejad2011analyzing} to incorporate prior information about the partial sparsity level of a given partitioning of the elements in sparse signals.
%
The authors apply a weighed $\ell_1$ minimization approach and find an upper-bound for the failure probability of the recovery method. The bound is derived based on the internal and external angles of a certain weighted polytope. The weights could then be optimized to minimize the resulting upper-bound. Unfortunately, this approach does not yield explicit expressions for the weights; in addition, since the probability bound is not necessarily tight, the optimality of such weights is questionable. As mentioned by the authors, the extension of this approach to block-sparse signals is very sophisticated.

Using a different approach, \cite{friedlander2012recovering} investigates the same model with weighted $\ell_1$ minimization. Indeed, the approach consists of dividing the signal elements into two subsets; the elements of the first set are penalized with a variable weight, while the weights of the second set are kept fixed at $1$. If the first set encompasses at least $50\%$ of the signal support, it is shown in 
\cite{friedlander2012recovering} that the weighting technique can improve the recovery performance.
%

The bipartitioning approach of \cite{friedlander2012recovering} can be considered as having two estimates of the support. The extension to multiple estimates for the support with varying degrees of accuracy is studied in \cite{needell2017weighted}. It is again confirmed that a multi-level weighting scheme can enhance the recovery performance. The tuning of the weights in \cite{needell2017weighted} is achieved based on empirical results rather than theoretical findings.


A systematic approach for determining the optimal weights is presented in \cite{flinth2016optimal}. Indeed, the optimal weights are derived by minimizing the number of required measurements for almost sure sparse recovery of a weighted $\ell_1$ minimization. The resulting optimal weights are expressed in terms of the accuracy level (normalized partial sparsity level) of the sets. Our paper could be considered as an extension of \cite{flinth2016optimal} to the block-sparse setting. However, our approach to derive theoretical results is substantially different from \cite{flinth2016optimal}.
For instance, deriving the optimal weights (and proving their uniqueness) for the $\mathsf{P}_{1,2,\bm{w}}^{\eta}$ problem is much more challenging than the weighted $\ell_1$ minimization.
Moreover, the construction of a tight upper-bound for the sample complexity that leads to closed-form expressions is nontrivial.

In many practical scenarios, the low-dimensional signal of interest has block-sparse structure rather than simple sparse structure (see Section \ref{sec.app} and \cite[Section I]{khajehnejad2011analyzing} for illustrative examples). In this work, we consider a non-uniform block-sparse model where we have a partitioning of the blocks and their associated accuracy levels.
This model extends the non-uniform sparse model considered in \cite{friedlander2012recovering, flinth2016optimal} to the non-uniform block-sparse signals which is observed in a few works such as \cite{babacan2014bayesian,som2010approximate,ahmad2015iteratively}. A distinctive feature of our work, besides the generality of our model, is the closed-form expressions for the optimal weights in the $\mathsf{P}_{1,2,\bm{w}}^{\eta}$ problem. In contrast to the result in \cite{friedlander2012recovering},
we show that the minimum $50\%$ accuracy level is not a pre-requisite for  $\mathsf{P}_{1,2,\bm{w}}^{0}$ to outperform the unweighted $\mathsf{P}_{1,2}^{0}$ recovery approach. We study the robustness of the weights under inaccuracies in prior knowledge for the first time, and prove robustness for the introduced optimal weights. The study of robustness has not been accomplished in the past even for the easier sparse models.
\subsection{DOA estimation}\label{sec.app}
DOA estimation is the problem of finding the direction of a few sources from observations on an array of sensors. The location of each source is characterized by the direction of arrival $\theta\in[-\frac{\pi}{2},\frac{\pi}{2}]$ with respect to the array axis. 
Let $\theta_i:=\pi(\tfrac{i-1}{q-1}-\tfrac{1}{2}), i=1,..., q$ be a discretization of the interval $[-\frac{\pi}{2},\frac{\pi}{2}]$ into $q$ grid points representing the potential direction of the sources (see the right block of Figure \ref{fig.model}). Assume that $s\ll q$ far-field narrow-band sources with wavelength $\lambda$ whose directions exactly match the grid angles incident on a $q$-element uniform linear array (ULA) of sensors with inter-sensor spacing $d$. Form the so-called array manifold matrix
$\bm{F}=[\bm{a}(\theta_1),..., \bm{a}(\theta_q)]\in\mathbb{C}^{q\times q}$ where $\bm{a}(\theta_l):={\rm e}^{{\rm j}2\pi (d/\lambda)[0,..., q-1]^T\sin(\theta_l)}$ is the steering vector caused by the propagation delay from $i$th potential source to each of the $q$ sensors. The observed signal at ULA elements at time $t$ are given by
\begin{align}\label{eq.doaeq1}
\bm{y}(t)=\bm{F}\,\bm{x}(t)+\bm{e}(t),
\end{align}
where $x_i(t)=\sqrt{p_i}{\rm e}^{{\rm j}2\pi i t}$ is the amplitude of the $i$th potential source, $\bm{y}(t)\in\mathbb{C}^q$ is the received vector at time (or snapshot) $t$ at all sensors, and $\bm{e}(t)$ is a Gaussian noise term with variance $\sigma_{e}^2$. In practice, the number of ULA sensors is limited due to cost limitations or physical constraints; hence, it is crucial in DOA estimation to exploit time redundancies (taking several snapshots, \emph{i.e.}, $\bm{y}(t_1)$, $\dots$, $\bm{y}(t_k)$). 
With the assumption that the sources remain fixed in all snapshots, the received signals at all $k$ time instances could be expressed as
\begin{align}
\bm{Y}:=[\bm{y}(t_1),..., \bm{y}(t_k)]=\bm{F}\bm{X}+\bm{E},
\end{align} 
where $\bm{X}=[\bm{x}(t_1),..., \bm{x}(t_k)]\in\mathbb{C}^{q\times k}$ and $\bm{E}$ is similarly defined. 
In practical scenarios, we can not directly observe the received signal at all of the sensors. Instead, we have access to the outputs of $m\ll q$ sensors formed by
\begin{align}\label{eq.doamodel}
\bm{Y}=\bm{A}\bm{F}\bm{X}+\widetilde{\bm{E}}\in\mathbb{C}^{m\times k},
\end{align}
where $\bm{A}\in\mathbb{C}^{m\times q}$ is the partial identity matrix \footnote{This means that only $m$ random rows of the identity matrix $\bm{I}_{q\times q}$ are chosen.}, $\bm{A F}$ represents the effective measurement matrix and $\widetilde{\bm{E}}:=\bm{A}\bm{E}\in\mathbb{C}^{m\times k}$.
In practical settings, the number of sources are limited. Thus, $\bm{X}$ consists mainly of zeros. In addition, if $\bm{X}$ is vectorized by concatenating the rows, we achieve a block-sparse vector. Indeed, each row forms a block in the vectorized signal. It is common in radar applications that the operator has prior information (previous measurement of recorded history) about the partial number of sources in a given angular partitioning (see $\{\mathcal{P}_i\}_{i=1}^5$ in the right block of Figure \ref{fig.model}). More precisely, the radar operator might know the accuracy ($\alpha_i$s in the right block of Figure \ref{fig.model}) of specific angular bands.
%
It is of considerable importance to exploit these information in order to minimize the number of required sensors for finding the sources. We will turn back to this problem in Section \ref{section.simulation}.

The paper is organized as follows: some concepts from convex geometry are reviewed in Section \ref{sec.prilim}. The signal model and our methodology are stated in Section \ref{section.model}. The number of measurements required for $\mathsf{P}_{1,2,\bm{w}}^{0}$ is obtained in Section \ref{section.measurement}. In Section \ref{section.optweights}, the procedure of finding optimal weights is explained. In Section \ref{section.robust}, the robustness of optimal weights with respect to inaccuracies in prior information is discussed. Numerical simulations on synthetic data that support the theory are presented in Section \ref{section.simulation}. Finally, the paper is concluded in Section \ref{section.conclusion}.

\textit{Notation} Throughout the paper, scalars are denoted by lowercase letters, vectors by lowercase boldface letters, and matrices by uppercase boldface letters. The $i$th element of a vector $\bm{x}$ is shown either by ${x}(i)$ or $x_i$. $[n]$ refers to $\{1,..., n\}$. We reserve calligraphic uppercase letters for sets (e.g. $\mathcal{S}$). The cardinality of a set $\mathcal{S}$ is shown by $|\mathcal{S}|$. ${\overline{\mathcal{S}}}$ denotes the complement $[n]\setminus\mathcal{
	S}$ of a set $\mathcal{S}\subset [n]$. For $\bm{x}\in\mathbb{R}^n$, $\bm{x}_\mathcal{S}$ is the subvector in $\mathbb{R}^{|\mathcal{S}|}$ consisting of the entries indexed by $\mathcal{S}$, that is, $(\bm{x}_S)_i = x_{j_i}~:~\mathcal{S}=\{j_i\}_{i=1}^{|\mathcal{S}|}$. In this paper, $\bm{1}_{\mathcal{E}}$ denotes the indicator of a set $\mathcal{E}$. The null space of linear operators is denoted by $\mathrm{null}(\cdot)$. Given a vector $\bm{x}\in\mathbb{R}^n$ and a set $\mathcal{C}\subseteq \mathbb{R}^n$, the set obtained by scaling elements of $\mathcal{C}$ by the elements of $\bm{x}$ is denoted by $\bm{x}\odot \mathcal{C}$. The polar $\mathcal{K}^{\circ}$ of a cone $\mathcal{K}\subset\mathbb{R}^n$ is the set of vectors forming non-acute angles with every vector in $\mathcal{K}$, i.e. 
$\mathcal{K}^\circ=\{\bm{v}\in\mathbb{R}^n: \langle \bm{v}, \bm{z} \rangle\le 0~\forall \bm{z}\in\mathcal{K}\}.$ For a matrix $\bm{A}$, the operator norm is defined as $\|\bm{A}\|_{p\rightarrow q}=\underset{\|\bm{x}\|_p\le1}{\sup}\|\bm{Ax}\|_q$. For $\bm{x},\bm{y}\in\mathbb{R}^n$, $\bm{x}\le \bm{y}$ stands for component-wise inequality while $\bm{x}<\bm{y}$ denotes component-wise inequality with strict inequality in at least one component. $\mathds{B}_{\epsilon}^n$ refers to the $\epsilon$-ball $\mathds{B}_{\epsilon}^n=\{\bm{x}\in \mathbb{R}^n:~\|\bm{x}\|_2\le\epsilon\}$. Lastly, the notation $(a)_+$ stands for the positive part of $a$, i.e.,  $\max\{a,0\}$.
\section{Preliminaries}\label{sec.prilim}
In this section, basic concepts of conic integral geometry are reviewed.

\subsection{Subdifferential}
The subdifferential of a proper\footnote{An everywhere defined function taking values in $(-\infty,\infty]$ with at least one finite value in $(-\infty,\infty)$.} convex function $f:\mathbb{R}^n\rightarrow \mathbb{R}\cup \{\pm\infty\}$ at $\bm{x}\in\mathbb{R}^n$ is given by:
\begin{align}\label{eq.subdiff}
\partial f(\bm{x}):=\{\bm{z}\in \mathbb{R}^n: f(\bm{y})\ge f(\bm{x})+\langle \bm{z},\bm{y}-\bm{x} \rangle~:~\forall \bm{y}\in \mathbb{R}^n\}.
\end{align}
\begin{prop}\label{prop.simplerform of subdiff}
	Let $f: \mathbb{R}^n\rightarrow\mathbb{R}\cup \{\pm\infty\}$ be a proper convex function that is $1$-homogeneous, i.e. $f(\alpha \bm{z})=|\alpha|f(\bm{z})~:\forall \alpha \in \mathbb{R}$ and sub-additive, i.e. $f(\bm{x}+\bm{y})\le f(\bm{x})+f(\bm{y})~~:\forall \bm{x},\bm{y} \in \mathbb{R}^n$, then we have a simpler form of subdifferential given by:
	\begin{align}\label{eq.subdiff2}
	\partial f(\bm{x})=
	&\small{\left\{\bm{z}\in\mathbb{R}^n:\begin{array}{lr}
		\langle \bm{z},\bm{x} \rangle=f(\bm{x}) , f^*(\bm{z})=1, &  \bm{x}\neq \bm{0}\\
		f^*(\bm{z})\le 1, & \bm{x}=\bm{0}
		\end{array}\right\}}\nonumber\\
	\end{align}
	where $f^*(\bm{z})=\underset{f(\bm{y})\le 1}{\sup}\langle \bm{z},\bm{y} \rangle$ is the dual function of $f(\bm{z})$.
\end{prop}
Proof. See Appendix \ref{proof.subdiff_simpler}.
\subsection{Descent Cones}
The descent cone of a proper convex function $f:\mathbb{R}^n\rightarrow \mathbb{R}\cup \{\pm\infty\}$ at point $\bm{x}\in \mathbb{R}^n$ is the conic hull of the directions from $\bm{x}$ that $f$ does not increase:
\begin{align}\label{eq.descent cone}
\mathcal{D}(f,\bm{x})=\bigcup_{t\ge0}\{\bm{z}\in\mathbb{R}^n: f(\bm{x}+t\bm{z})\le f(\bm{x})\}\cdot
\end{align}
The descent cone of a convex function is a convex set. There is also a relation between decent cone and subdifferential of a convex function\cite[Chapter 23]{rockafellar2015convex} which is given by:
\begin{align}\label{eq.D(f,x)}
\mathcal{D}^{\circ}(f,\bm{x})=\mathrm{cone}(\partial f(\bm{x})):=\bigcup_{t\ge0}t.\partial f(\bm{x}).
\end{align}

\subsection{Statistical Dimension}
\begin{defn}{Statistical Dimension}\cite{amelunxen2013living}:
	Let $\mathcal{C}\subseteq\mathbb{R}^n$ be a convex closed cone. Statistical dimension of $\mathcal{C}$ is defined as:
	\begin{align}\label{eq.statisticaldimension}
	\delta(\mathcal{C}):=\mathds{E}_{\bm{g}}\|\mathcal{P}_\mathcal{C}(\bm{g})\|_2^2=\mathds{E}_{\bm{g}}\mathrm{dist}^2(\bm{g},\mathcal{C}^\circ),
	\end{align}
	where
	\begin{align}
	 \mathcal{P}_\mathcal{C}(\bm{x}):=\underset{\bm{z} \in \mathcal{C}}{\arg\min}\|\bm{z}-\bm{x}\|_2,
	 \end{align}
	is the projection of $\bm{x}\in \mathbb{R}^n$ onto the set $\mathcal{C}$, and $\bm{g}$ is a vector with i.i.d. elements each following a standard normal distribution.
\end{defn}
Statistical dimension is a measure of the size of a convex cone and generalizes the concept of dimension for linear subspaces to the class of convex cones.
\subsection{Inverse Problems via Convex Optimization}
Convex optimization is a common approach for recovering a structured signal $\bm{x}_{n\times 1}$ from $m$ linear measurements of the form  \eqref{eq.linear_measure}. Let $f(\cdot)$ be a function that promotes the structure of $\bm{x}$ (e.g., sparsity). Then, one might be able to recover the signal $\bm{x}$ by solving the problem
\begin{align}\label{eq.mainprob}
\mathsf{P}_f^{\eta}:~~~\min_{\bm{z}\in \mathbb{R}^n} ~f(\bm{z})~~~~
\mathrm{s.t.} ~~\|\bm{Az}-\bm{y}\|_2\le \eta,
\end{align}
where $\eta$ is a known upper-bound on the norm of the noise $\bm{e}$. For the noiseless case of $\bm{e}=\bm{0}$ in \eqref{eq.linear_measure}, it is shown in \cite{amelunxen2013living} that $\mathsf{P}_f^{0}$ exhibits a sharp phase-transition behavior (success/failure) as the number of measurements increases. 
In addition, the boundary of the transition is determined by the statistical dimension of the decent cone of $f$ at $\bm{x}$, i.e. $\delta(\mathcal{D}(f,\bm{x}))$.

In Theorem \ref{thm.Pfmeasurement}, we present a result regarding the recovery performance of \eqref{eq.mainprob}
from random measurements, in both noiseless and noisy settings. This theorem is an adaptation  of \cite[Corollary 3.5]{tropp2015convex} and \cite[Theorem 2]{amelunxen2013living}.

\begin{thm}\label{thm.Pfmeasurement}
	Let $f(\cdot)$ be a proper convex function that promotes the structure of $\bm{x}$. Let $\bm{A}_{m\times n}$ be a random matrix whose null space is uniformly distributed with respect to the Haar measure. Then, if
	\begin{align}
	m\ge \delta(\mathcal{D}(f,\bm{x}))+\sqrt{8\log(\tfrac{4}{\zeta})n}\nonumber
	\end{align}
	for some $\zeta \in [0,1]$, then $\mathsf{P}_f^{0}$ recovers $\bm{x}$ from $\bm{y}_{m\times 1}=\bm{A x}$ with probability at least $1-\zeta$. Alternatively, in the noisy case of $\bm{y}_{m\times 1}=\bm{A x}+\bm{e}$, where $\|\bm{e}\|_2\le \eta$, if  $\widehat{\bm{x}}_{\eta}$ is any solution of $\mathsf{P}_f^{\eta}$, then
	\begin{align}\label{eq.recovery_error}
	\|\widehat{\bm{x}}_{\eta}-\bm{x}\|_2\le \tfrac{2\eta}{(\sqrt{m-1}-\sqrt{\delta(\mathcal{D}(f,\bm{x}))}-\zeta)_{+}},
	\end{align} 
	with probability at least $1-{\rm e}^{-\tfrac{\zeta^2}{2}}$.
\end{thm} 
Also in \cite{amelunxen2013living}, the following error bound for the statistical dimension is provided.
\begin{thm}\cite[Theorem 4.3]{amelunxen2013living} For any $\bm{x}\in \mathbb{R}^n\setminus\{\bm{0}\}$:
	\begin{align}\label{eq.errorbound}
	0\le\inf_{t\ge0}\mathds{E}\mathrm{dist}^2(\bm{g},t\partial f(\bm{x}))- \delta(\mathcal{D}(f,\bm{x}))\le \tfrac{2\sup_{s\in \partial f(\bm{x})}\|s\|_2}{f(\tfrac{\bm{x}}{\|\bm{x}\|_2})}.
	\end{align}
\end{thm}
\section{Model and methodology}\label{section.model}
Suppose $\bm{x}_{n\times 1}$ is a $s$-block-sparse signal consisting of $q$ blocks $\{\mathcal{V}_b\}_{b=1}^q$ of equal size $k$, among which only $s$ blocks are non-zero. We are further given a partition of $\{1,\dots, q\}$ into $L$ disjoint subsets $\mathcal{P}_i\subseteq \{1,\dots q\}$, for which we know
%
%
\begin{align}
\alpha_i=\tfrac{|\mathcal{P}_i\cap\mathcal{B}|}{|\mathcal{P}_i|},~\rho_i=\tfrac{|\mathcal{P}_i|}{q}, i=1,..., L,
\end{align}
where $\mathcal{B}\subseteq\{1,\dots,q\}$ is the block-support of  $\bm{x}_{n\times 1}$.
Intuitively, $\alpha_i$ (which is called the accuracy of $\mathcal{P}_i$) stands for the normalized block-sparsity level of $\bm{x}$ restricted to $\mathcal{P}_i$.
%
We call $\bm{x}$ a non-uniform block-sparse model with parameters $\{\alpha_i\}_{i=1}^L$ and $\{\rho_i\}_{i=1}^L$; if $\alpha_i$'s are all equal, the non-uniform model $\mathsf{P}_{1,2,\bm{w}}^{\eta}$ reduces to the uniform model $\mathsf{P}_{1,2}^{\eta}$. Each set $\mathcal{P}_i$ is associated with a weight $\omega_i\ge0$ and the resulting weight $\bm{w}$ in $\mathsf{P}_{1,2,\bm{w}}^{\eta}$ is 
\begin{align}
\bm{w}=\bm{D}\bm{\omega},
\end{align}
where $\bm{D}:=[\bm{1}_{\mathcal{P}_1},...,\bm{1}_{\mathcal{P}_L}]\in\mathbb{R}^{q\times L}$. To better distinguish between $\omega_i$ and $w_i$, note that the former penalizes the index set $\mathcal{P}_i$ while the latter penalizes the index set $\mathcal{V}_i$ (see the illustrations in the left image of Figure \ref{fig.model}). 

In this work, the following questions are answered about a non-uniform block-sparse model:
\begin{enumerate}
  \item How many measurements are required for $\mathsf{P}_{1,2}^{0}$ and $\mathsf{P}_{1,2,\bm{w}}^{0}$ to successfully recover
a $s$-block-sparse vector from independent linear measurements?
  \item Given extra prior information, what is the optimal choice of weights in $\mathsf{P}_{1,2,\bm{w}}^{\eta}$?
  \item How close one can get to the optimal weights if the prior information is slightly inaccurate?
\end{enumerate}
In the reminder of this work, we provide the answer to these questions in three sections.

\section{Number of Measurements for Successful Recovery}\label{section.measurement}
For a fixed tolerance $\zeta\in [0,1]$, let us denote the normalized number of measurements required for $\mathsf{P}_{1,2}^{0}$ and $\mathsf{P}_{1,2,\bm{w}}^{0}$ to recover a $s$-block-sparse vector (with probability $1-\zeta$) by $m_{q,s}$ and $m_{q,s,\bm{w}}$, respectively:
\begin{align}
m_{q,s}:=\tfrac{\delta(\mathcal{D}(\|\cdot\|_{1,2},\bm{x}))}{q},~~m_{q,s,\bm{w}}:=\tfrac{\delta(\mathcal{D}(\|\cdot\|_{1,2,\bm{w}},\bm{x}))}{q}.
\end{align}
Below, we obtain an upper-bound for the number of measurements required for $\mathsf{P}_{1,2}^{0}$ to succeed with probability $1-\zeta$.
\begin{lem}\label{lemma.mhat qsw}
	Let $\bm{x}\in \mathbb{R}^n$ be a non-uniform $s$-block-sparse vector in $\mathbb{R}^{n}$ with parameters $\{\rho_i\}_{i=1}^L$ and $\{\alpha_i\}_{i=1}^L$. Then,
	\begin{align}
	m_{q,s,\bm{w}}\le \widehat{m}_{q,s,\bm{w}}
	\end{align}
	for
	\begin{align}\label{eq.hatm-qsw}
	\widehat{m}_{q,s,\bm{w}}=\underset{{t\ge0}}{\inf}~\Psi_{t,\bm{w}}(\sigma,\bm{\rho},\bm{\alpha}),
	\end{align}
	where
	\begin{align}
	&\Psi_{t,\bm{w}}(\sigma,\bm{\rho},\bm{\alpha})=
	\sum_{i=1}^{L}\rho_i\big(\alpha_i(k+t^2\omega_i^2)+\tfrac{(1-\alpha_i)\phi(t\omega_i)}{2^{\tfrac{k}{2}-1}\Gamma(\tfrac{k}{2})}\big),\nonumber\\
	&\phi(z):=\int_{z}^{\infty}(u-z)^2u^{k-1}\exp(-\tfrac{u^2}{2}){\rm{d}}u,~k=\tfrac{n}{q}.
	\end{align}
\end{lem}
Proof. See Appendix \ref{proof.lemma.mhat qsw}.
\begin{corl}\label{corl.mhat_qs}
By considering $\bm{w}=\bm{1}\in\mathbb{R}^q$, and using the fact that the normalized block-sparsity level is $\sigma:=\tfrac{\|\bm{x}\|_{0,2}}{q}=\sum_{i=1}^L\rho_i\alpha_i$, we reach an upper-bound $\widehat{m}_{q,s}$ for $m_{q,s}$  as
\begin{align}\label{eq.mhat_qs}
\widehat{m}_{q,s}=\underset{{t\ge0}}{\inf}~\Psi_t(\sigma),
\end{align}
where
\begin{align}
&\Psi_t(\sigma)=\sigma(k+t^2)+\tfrac{(1-\sigma)\phi(t)}{2^{\tfrac{k}{2}-1}\Gamma(\tfrac{k}{2})}.\nonumber
\end{align}
\end{corl}
\begin{rem}(Prior work) 
In \cite[Lemma 3.2]{donoho2013accurate}, the same expression as for $\widehat{m}_{q,s}$ is obtained for the normalized minimax MSE of the denoising problem
\begin{align}
\min_{\bm{z}\in \mathbb{R}^n}\tau\|\bm{z}\|_{1,2}+\tfrac{1}{2}\|\bm{y}-\bm{z}\|_2^2,
\end{align}
where $\bm{y}=\bm{x}+\bm{e}$ is the observed noisy vector. It is further conjectured in \cite{donoho2013accurate} that this value is equal to the number of measurements required by $\mathsf{P}_{1,2}^{0}$ in the asymptotic regime. 
%
We show that this formula describes the required number of measurements in $\mathsf{P}_{1,2}^{0}$ even in the non-asymptotic case (Proposition \ref{prop.error for mhat qsw}).
\end{rem}

%
%
%
In the following Proposition, we demonstrate that the proposed upper-bound in  Lemma \ref{lemma.mhat qsw} is asymptotically tight. We use this fact for the optimality of the obtained weights.

\begin{prop}\label{prop.error for mhat qsw}
The normalized number of linear measurements required for $\mathsf{P}_{1,2,\bm{w}}^{0}$ and $\mathsf{P}_{1,2}^{0}$ to successfully recover a non-uniform $s$-block-sparse vector in $\mathbb{R}^n$ (i.e. $m_{q,s,\bm{w}}$ and $m_{q,s}$, respectively ) satisfy the following error bounds:
\begin{align}
&\widehat{m}_{q,s,\bm{w}}-\tfrac{2}{\sqrt{qL}}\le m_{q,s,\bm{w}}\le \widehat{m}_{q,s,\bm{w}},\label{eq.l12w error}\\
&\widehat{m}_{q,s}-\tfrac{2}{\sqrt{sq}}\le m_{q,s}\le \widehat{m}_{q,s}\label{eq.l12 error}.
\end{align}
\end{prop}
Proof. See Appendix \ref{proof.mhatqsw}.

It is interesting that the error bound in \eqref{eq.l12 error} is a special case of the error bound of Proposition \eqref{eq.l12w error} where one has $s$ sets of blocks with size $\tfrac{q}{L}$ that each contributes to the block support with probability $\tfrac{L}{q}$.
\section{Optimal Weights}\label{section.optweights}
Our strategy in finding the optimal weights is to minimize the reconstruction error \eqref{eq.recovery_error} in the noisy case (under a fixed number of measurements i.e. $m$), and the required number of measurements in the noiseless case. Based on Theorem \ref{thm.Pfmeasurement}, both of these objectives lead to the same optimization problem
\begin{align}
\bm{\omega}^*=\underset{\bm{\omega}\in\mathbb{R}_{+}^L}{\arg\min}~~m_{q,s,\bm{D\omega}}\in\mathbb{R}_{+}^L.
\end{align}
Instead of the latter minimization, we minimize the upper and lower bounds of statistical dimension (i.e. \eqref{eq.l12w error}), simultaneously. So
\begin{align}\label{eq.infmhatqserrorbound}
\inf_{\bm{\omega}\in\mathbb{R}_{+}^L}\widehat{m}_{q,s,\bm{D\omega}}-\tfrac{2}{\sqrt{qL}}\le\inf_{\bm{\omega}\in\mathbb{R}_{+}^L}{m}_{q,s,\bm{D\omega}}\le \inf_{\bm{\omega}\in\mathbb{R}_{+}^L}\widehat{m}_{q,s,\bm{D\omega}},
\end{align}
where $\bm{D}:=[\bm{1}_{\mathcal{P}_1},..., \bm{1}_{\mathcal{P}_L}]_{q\times L}$.
In the weighted block sparsity optimization, we call the weight
\begin{align}\label{eq.opt_weight_cal}
\bm{\omega}^*=\underset{\bm{\omega}\in\mathbb{R}_{+}^L}{\arg\min}~~\widehat{m}_{q,s,\bm{D\omega}}\in\mathbb{R}_{+}^L
\end{align}
  optimal since it asymptotically (as $q\rightarrow \infty$) minimizes simultaneously the number of measurements required for $\mathsf{P}_{1,2,\bm{w}}^{0}$ to succeed, and the reconstruction error of $\mathsf{P}_{1,2,\bm{w}}^{\eta}$. In the following lemma, the uniqueness of the optimal weights 
  is shown by proving that $\delta(\mathcal{D}(\|\cdot\|_{1,2,\bm{D}\bm{\omega}},\bm{x}))$ is a strictly convex function of $\bm{\omega}\in\mathbb{R}_{+}^L$.
\begin{lem}{\label{lemma.Jb(nu)}}
Assume $\mathcal{C}:=\partial\|\cdot\|_{1,2}(\bm{x})$ does not contain the origin. We know that $\mathcal{C}$ is compact and $1\le\|\bm{z}\|_2\le \sqrt{q}$ for all $\bm{z}\in \mathcal{C}$. Also let $\bm{g}\in \mathbb{R}^n$ be a standard normal vector. Consider the function
\begin{align}
J(\bm{\nu}):=\mathds{E}_{\bm{g}}\mathrm{dist}^2(\bm{g},\bm{\upsilon}\odot \mathcal{C})=\mathds{E}_{\bm{g}}[J_{\bm{g}}(\bm{\nu})] \nonumber\\
 \text{with}~~\bm{\upsilon}=\bm{D}_b\bm{D}\bm{\nu} \in \mathbb{R}^n ~~\text{for}~~ \bm{\nu}\in\mathbb{R}_{+}^L,
\end{align}
where $\bm{D}_b:=[\bm{1}_{\mathcal{V}_1},..., \bm{1}_{\mathcal{V}_q}]_{n\times q}$. Then, the function $J$ is strictly convex on $\bm{\nu}\in \mathbb{R}_{++}^L$ and $J$ has a unique minimizer.
\end{lem}
Proof. See Appendix \ref{proof.lem_convex}.

An analytic expression for the optimal weights is given in the following proposition via solving \eqref{eq.opt_weight_cal}. 
\begin{prop}\label{prop.uniqnessoptimal12weights}
Let $\bm{x}$ be a non-uniform $s$-block-sparse vector in $\mathbb{R}^n$ with parameters $\{\rho_i\}_{i=1}^L$ and $\{\alpha_i\}_{i=1}^L$. Then, there exist unique optimal weights $\bm{\omega}^* \in \mathbb{R}_{+}^L$ (up to a positive scaling) that minimize $\widehat{m}_{q,s,\bm{D\omega}}$. The optimal weights $\bm{\omega}^*\in\mathbb{R}_{+}^L$ are obtained via the following integral equations:
\begin{align}\label{eq.12optimalweights}
&\alpha_i\omega^*_i=\tfrac{1}{2^{\tfrac{k}{2}-1}\Gamma(\tfrac{k}{2})}(1-\alpha_i)\int_{\omega^*_i}^{\infty}(u-\omega^*_i)u^{k-1}{\rm{e}}^{-\tfrac{u^2}{2}}{\rm{d}}u\nonumber\\
&i=1,..., L.
\end{align}
\end{prop}
Proof. See Appendix \ref{proof.uniqnessoptimal12weights}\\
The optimal weights in (\ref{eq.12optimalweights}) depend only on the accuracy of $\mathcal{P}_i$s, i.e., $\{\alpha_i\}_{i=1}^L$ and not their relative size $\{\rho_i\}_{i=1}^L$. Finally, we prove the significant fact that with optimal weights, $\mathsf{P}_{1,2,\bm{D}\bm{\omega}^*}^{0}$ acts as if $\bm{x}_{\mathcal{P}_i}$'s are separately recovered via $\mathsf{P}_{1,2}^{0}$, in the sense of the required number of measurements. 
\begin{thm}\label{thm.l12weighted}
Let $\bm{x}$ be a non-uniform $s$-block-sparse vector in $\mathbb{R}^{n}$ with parameters $\{\rho_i\}_{i=1}^L$ and $\{\alpha_i\}_{i=1}^L$. Then, the number of measurements required for $\mathsf{P}_{1,2,\bm{D\omega}^*}^{0}$ is exactly equals the total number of measurements required for $\mathsf{P}_{1,2}^{0}$ to recover each $\{\bm{x}_{\mathcal{P}_i}\in\mathbb{R}^n\}_{i=1}^L$ separately, up to an asymptotically additive vanishing error term i.e.
\begin{align}\label{eq.rel1}
&-\tfrac{2}{\sqrt{qL}}\le m_{q,s,\bm{D\omega}^*}-\sum_{i=1}^{L}m_{q,\|\bm{x}_{\mathcal{P}_i}\|_{0,2}}
\le\tfrac{2}{\sqrt{q}}\sum_{i=1}^{L}(\|\bm{x}_{\mathcal{P}_i}\|_{0,2})^{-\tfrac{1}{2}}.
\end{align}
\end{thm}
\begin{figure}[t]
	\hspace{-0.5cm}
	\includegraphics[scale=.17]{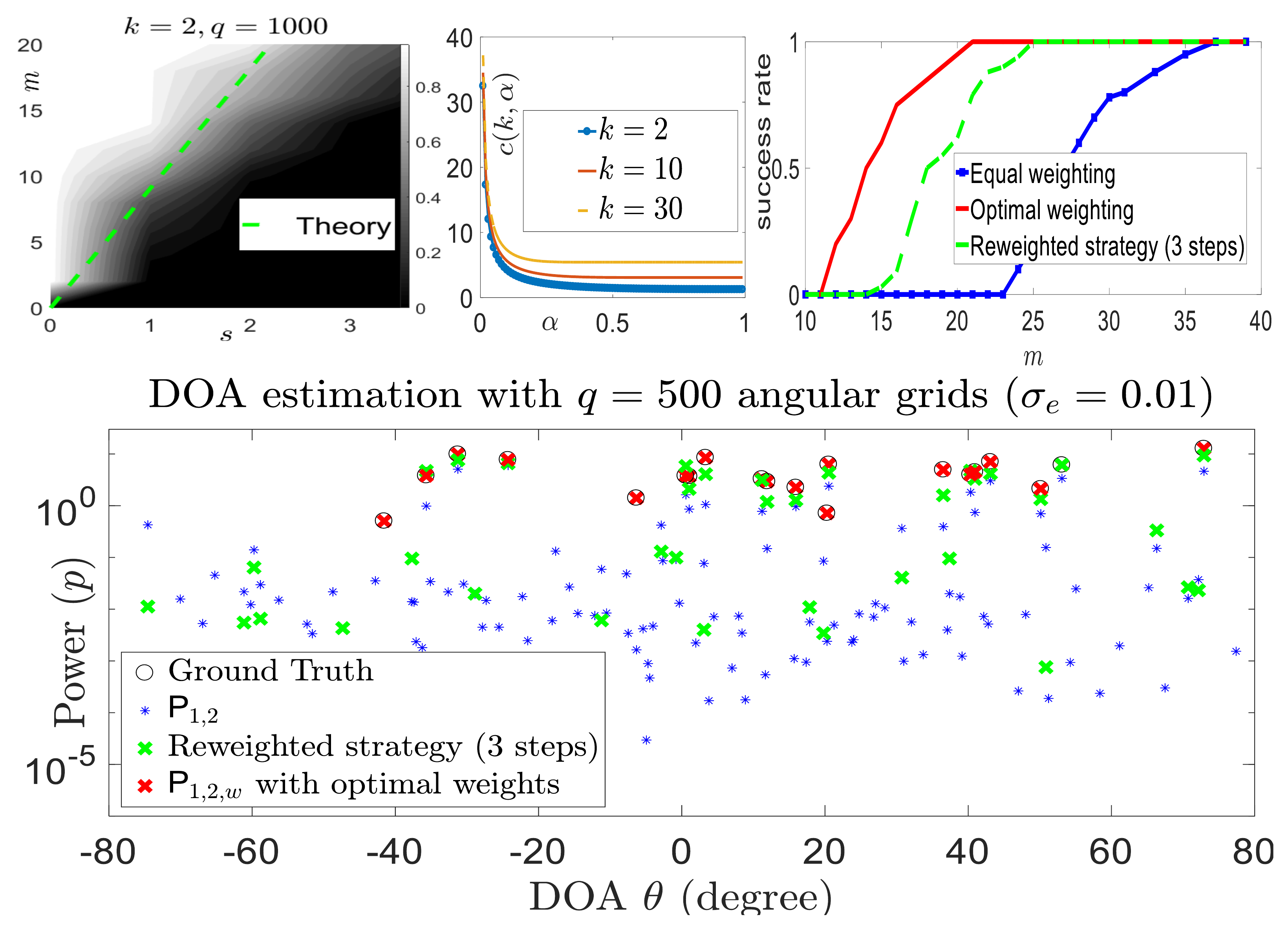}
	\caption{Top left image: This plot shows the empirical probability that $\mathsf{P}_{1,2}^{0}$ recovers $\bm{x}\in\mathbb{R}^{2000}$ with $s$ blocks with nonzero $\ell_2$ norm from $m$ Gaussian linear measurements. The green dashed line shows the number of measurements obtained by Corollary \ref{corl.mhat_qs}. Top middle image: This plot shows $c(k,\alpha)$ in (\ref{eq.c(k)}) versus the accuracy $\alpha$ for $k=2, 10, 30$. Indeed, it demonstrates the sensitivity of the optimal weight against small perturbations of $\alpha$. Top right image: This plot shows the probability that $\mathsf{P}_{1,2,\bm{w}}^{0}$ succeed to recover $\bm{x}\in\mathbb{R}^{1280}$ from Gaussian linear measurements when $\bm{w}$ is chosen equally, optimally and via a reweighted strategy (3 steps) proposed in \cite{ahmad2015iteratively}. The parameters we used in this figure, are: $q=128$, $\sigma=\rho_1=\tfrac{10}{128},~\alpha_1=\tfrac{9}{10}$, $\rho_2=\tfrac{118}{128},~\alpha_2=\tfrac{1}{118}$. The optimal weights obtained via (\ref{eq.12optimalweights}) with the aforementioned parameters are $\bm{\omega}^*=[.0766,1]^T$. Bottom image: DOA estimation of $s=20$ sources in the angular half-space $[-\frac{\pi}{2},\frac{\pi}{2}]$ with $k=10$ snapshots, $m=24$, $q=500$ and $\tfrac{d}{\lambda}=1$. The measurements are contaminated with additive Gaussian noise with $\sigma_e=0.01$. Also, we consider four angular bands with accuracies $\alpha_1=\tfrac{5}{6}$, $\alpha_2=\tfrac{13}{14}$, $\alpha_3=\tfrac{2}{3}$ and $\alpha_4=0$ as prior information.}
	\label{fig.p+s+w+r}
\end{figure}
\section{Robustness Analysis}\label{section.robust}
In this section, we evaluate the robustness of optimal weights if prior information is slightly inaccurate. As stated in the below proposition, under the condition $\alpha\gtrapprox\tfrac{1}{10}$, the optimal weights $\bm{\omega}^*$ in (\ref{eq.12optimalweights}) are robust to inaccuracies in the prior information.   
\begin{prop}\label{prop.robustness}
Let $\alpha$ be the accuracy of a set $\mathcal{P}$ which inaccurately assumed $\alpha'$ in practice. Let $\omega$ and $\omega'$ be the optimal weights, obtained from (\ref{eq.12optimalweights}), corresponding to $\alpha$ and $\alpha'$, respectively. Then, there exists a constant $c(k,\alpha)$ such that
\begin{align}
|\omega-\omega'|\le c(k,\alpha)|\alpha-\alpha'|
\end{align}
for 
\begin{align}\label{eq.c(k)}
c(k,\alpha):=\tfrac{\bigg(\sqrt{2}h(\alpha)\big(\Gamma(\tfrac{k}{2})-\gamma(\tfrac{k}{2},\tfrac{h(\alpha)^2}{2})\big)+2\gamma(\tfrac{k}{2},\tfrac{h(\alpha)^2}{2})\bigg)^2}{2\sqrt{2}\Gamma(\tfrac{k}{2})\gamma(\tfrac{k}{2},\tfrac{h(\alpha)^2}{2})},
\end{align}
where $\gamma(a,z):=\int_{z}^{\infty}u^{a-1}{\rm{e}}^{-u}{\rm{d}}u$ is incomplete gamma function and $h(\alpha)$ is the nonlinear function in (\ref{eq.12optimalweights}) that relates $\alpha$ to the optimal weight $\omega$.
\end{prop}
Proof. See Appendix \ref{proof.robust}.

This proposition shows that our method of finding optimal weights in (\ref{eq.12optimalweights}) is robust to inaccuracies in prior knowledge.

\section{Simulations}\label{section.simulation}
In this subsection, we numerically verify our theoretical results on the optimal weights. We have employed the CVX MATLAB package \cite{cvx} to implement optimization problems. First, we investigate the required number of measurements for the scaling of successful recovery of $\mathsf{P}_{1,2}^{0}$ in terms of the block sparsity. For this purpose, we construct a $s$-block-sparse $\bm{x}\in\mathbb{R}^{2000}$. Then, we form the measurements $\bm{y}_{m\times 1}=\bm{A}\bm{x}$ and obtain an estimate $\widehat{\bm{x}}$ by solving the problem $\mathsf{P}_{1,2}^{0}$; we repeat this experiment for $100$ realizations of $\bm{A}$. For each experiment, we declare success if $\|\bm{x}-\widehat{\bm{x}}\|_2\le 10^{-4}$. The heatmap in the top left image of Figure \ref{fig.p+s+w+r} shows the empirical probability of success for this procedure (black=$0\%$, white=$100\%$) which is consistent with the theory obtained by (\ref{eq.mhat_qs}).

In the second experiment, we set $s=10$ and generate a block-sparse random vector $\bm{x}\in\mathbb{R}^{1280}$ with $q=128$ blocks of equal size $k=10$. The block support of $\bm{x}$ is drawn uniformly at random and the values within each non-zero block are drawn from i.i.d. standard normal distribution. Then, we consider two sets $\mathcal{P}_1$, $\mathcal{P}_2$ with $\alpha_1=\tfrac{9}{10},~\alpha_2=\tfrac{1}{118}$ that partition the set of blocks $\{1,..., 128\}$. There are $100$ Monte Carlo trials for each $m$ where in each, we solve $\mathsf{P}_{1,2,\bm{D}\bm{\omega}}^{0}$ with optimal weights $\bm{\omega}^*$, equal weights and the reweighted strategy (3 steps) proposed in \cite{ahmad2015iteratively} and recover $\bm{x}\in\mathbb{R}^{1280}$ from $m$ Gaussian linear measurements. Note that $\bm{x}$ is kept fixed in Monte Carlo trials and only $\bm A$ changes. Optimal weights $\bm{w}^*$ are obtained from (\ref{eq.12optimalweights}) by MATLAB function \textsf{fzero}. The top right image of Figure \ref{fig.p+s+w+r} shows that $\mathsf{P}_{1,2,\bm{D}\bm{\omega}}^{0}$ with optimal weights needs fewer measurements than $\mathsf{P}_{1,2}^{0}$ and the reweighted method. To better investigate the sensitivity of optimal weights against small perturbations of $\alpha$, we have plotted the robustness constant $c(k,\alpha)$ (see (\ref{eq.c(k)})) in the top middle image of Figure \ref{fig.p+s+w+r}. These curves confirm the stability of optimal weights against the inaccuracy of prior information in the range $\alpha\gtrapprox\tfrac{1}{10}$.

In the last experiment, we investigate DOA estimation. We consider $s=20$ sources. As shown in the right block of Figure \ref{fig.model}, these sources are scattered in the angular half-space $[-\frac{\pi}{2},\frac{\pi}{2}]$ which is divided into $q$ angular grids. In practical scenarios, it is reasonable for an engineer to know the likelihood of source appearance in some known angular bands ($\alpha_i$ in our non-uniform block-sparse model). This information can be obtained for instance by considering the statistics of previous estimations or the characteristics of the underlying model. In this experiment, we consider four angular bands with accuracies $\alpha_1=\tfrac{5}{6}$, $\alpha_2=\tfrac{13}{14}$, $\alpha_3=\tfrac{2}{3}$ and $\alpha_4=0$. We have implemented $\mathsf{P}_{1,2,\bm{w}}^{\eta}$ using CVX software when $\bm{w}$ is chosen equally, optimally and via the reweighted strategy proposed in \cite{ahmad2015iteratively}. The optimal weights are calculated using (\ref{eq.12optimalweights}). We follow the model (\ref{eq.doamodel}) with $d=\lambda$, $q=500$, $\sigma_e=0.01$, and $m=24$. As it turns out from bottom image of Figure \ref{fig.p+s+w+r}, $\mathsf{P}_{1,2,\bm{w}^*}^{\eta}$ exactly recovers the direction and the power of the sources in contrast to $\mathsf{P}_{1,2}^{\eta}$ and the reweighted strategy which both fail. This in turn reveals the fact that, with optimal weighting strategy, much less sensors are required for stable recovery.
\section{Conclusion}\label{section.conclusion}
In this paper, we studied the recovery of block-sparse vectors using a weighted $\ell_{1,2}$-minimization, when some prior information about the distribution of the block-support is available. We have particularly studied the case where partial block-sparsity level of the vector within a block partitioning is available as prior information. Our goal was to find the optimal weights so as to minimize the number of required measurements for perfect recovery in the noiseless case, and to minimize the reconstruction error in the noisy case. In both cases, we simplified the task to minimizing the statistical dimension of a weighted descent cone. We introduced closed-form expressions that identify the unique optimal weights. We have also shown the robustness of the optimal weights against inaccuracies of the prior information.
\appendix
\section{Proof of Lemmas and Propositions}
\subsection{Proof of Proposition \ref{prop.simplerform of subdiff}}\label{proof.subdiff_simpler}
Assume first that $\bm{x}\neq\bm{0}$. By setting $\bm{y}=\bm{0}$ in (\ref{eq.subdiff}), we have:
	$\langle \bm{z},\bm{x} \rangle \ge f(\bm{x})-f(\bm{0})$ and $f(\bm{0})={0}$ due to the homogeneity, $f^*(\bm{z})\ge 1$. Also, setting $\bm{y}=\bm{v}+\bm{x}$ in (\ref{eq.subdiff}) under condition $f(\bm{v})=1$ we have:
	\begin{align}
	f^*(\bm{z})=\sup_{f(\bm{v})=1}\langle \bm{z},\bm{v} \rangle \le \sup_{f(\bm{v})=1}(f(\bm{v}+\bm{x})-f(\bm{x}))\le 1,
	\end{align}
	where we used sub-additivity of $f$. Hence, we have $f^*(\bm{z})=1$ and subsequently $\langle \bm{z},\bm{x} \rangle = f(\bm{x})$.\par
	On the other hand, if we have $\bm{z}$ such that $f^*(\bm{z})=1$ and $\langle \bm{z},\bm{x} \rangle = f(\bm{x})$, then for each $\bm{y}\in\mathbb{R}^n$:
	\begin{align}
	f(\bm{x})+\langle \bm{z},\bm{y}-\bm{x} \rangle = \langle \bm{z}, \bm{x}\rangle+\langle \bm{z},\bm{y}-\bm{x} \rangle \le f(\bm{y})\cdot
	\end{align}
	For $\bm{x}=\bm{0}$ in (\ref{eq.subdiff}), $\langle \bm{z},\bm{y} \rangle\le f(\bm{y})$ and then $f^*(\bm{z})\le 1$. Further, $f^*(\bm{z})\le 1$ implies that $\bm{z}\in \partial f(\bm{0})$.
\subsection{Proof of Lemma \ref{lemma.mhat qsw}}\label{proof.lemma.mhat qsw}
	By the definition of statistical dimension for $\mathcal{D}(\|\cdot\|_{1,2,\bm{w}},\bm{x})$ in (\ref{eq.statisticaldimension}), the fact that infimum of an affine function is concave and Jensen's inequality, we can find an upper-bound for $m_{q,s,\bm{w}}$ as:
	\begin{align}\label{eq.mhatqsw}
	m_{q,s,\bm{w}}\le \inf_{t\ge0}\underbrace{{q^{-1}}\mathds{E}_{\bm{g}}\mathrm{dist}^2(\bm{g},t\partial\|\cdot\|_{1,2,\bm{w}}( \bm{x}))}_{\Psi_{t,\bm{w}}(\sigma,\bm{\rho},\bm{\alpha})}:=\widehat{m}_{q,s,\bm{w}}.
	\end{align}
	The next step is to calculate $\partial\|\cdot\|_{1,2,\bm{w}}(\bm{x})$. From Proposition \ref{prop.simplerform of subdiff}, we have:
	\begin{align}\label{eq.subw12asli}
	\partial\|\cdot\|_{1,2,\bm{w}}(\bm{x})=\{\bm{z}\in \mathbb{R}^p:~\langle \bm{z}, \bm{x}\rangle=\| \bm{x}\|_{1,2,\bm{w}},~\|\bm{z}\|_{1,2,\bm{w}}^*=1\}.
	\end{align}
	It is not hard to show that,
	\begin{align}\label{eq.l12subdiff}
	\partial\|\cdot\|_{1,2,\bm{w}}(\bm{x}) = \left\{\bm{z}\in\mathbb{R}^n:\begin{array}{lr}
	\tfrac{w_b~\bm{x}_{\mathcal{V}_b}}{\|\bm{x}_{\mathcal{V}_b}\|_2}, &  b\in \mathcal{B}\\
	\|\bm{z}_{\mathcal{V}_b}\|_2\le w_b, & b\in {\overline{\mathcal{B}}}
	\end{array}\right\}.
	\end{align}
	Now to calculate $\Psi_{t,\bm{w}}(\sigma,\bm{\rho},\bm{\alpha})$, regarding (\ref{eq.l12subdiff}), we compute the distance of the dilated subdifferential of descent cone of $\ell_{1,2,\bm{w}}$ norm at $\bm{x}\in\mathbb{R}^n$ from a standard Gaussian vector $\bm{g}\in\mathbb{R}^n$ which is given by:
	\begin{align}\label{eq.distsubdiff12}
	&\mathrm{dist}^2(\bm{g},t\partial\|\cdot\|_{1,2,\bm{w}}(\bm{x}))=\inf_{\bm{z}\in\partial\|\cdot\|_{1,2,\bm{w}}(\bm{x})}\|\bm{g}-t\bm{z}\|_2^2=\nonumber\\
	&\sum_{b\in \mathcal{B}}{\|\bm{g}_{\mathcal{V}_b}-tw_b\tfrac{\bm{x}_{\mathcal{V}_b}}{\|\bm{x}_{\mathcal{V}_b}\|_2}\|_2^2}+\sum_{b\in {{\overline{\mathcal{B}}}}}\inf_{\|\bm{z}_{\mathcal{V}_b}\|_2\le w_b}{\|\bm{g}_{\mathcal{V}_b}-t\bm{z}_{\mathcal{V}_b}\|_2^2}=\nonumber\\
	&\sum_{b\in \mathcal{B}}{\|\bm{g}_{\mathcal{V}_b}-tw_b\tfrac{\bm{x}_{\mathcal{V}_b}}{\|\bm{x}_{\mathcal{V}_b}\|_2}\|_2^2}+\sum_{b\in {{\overline{\mathcal{B}}}}}{(\|\bm{g}_{\mathcal{V}_b}\|_2-tw_b)_{+}^2},
	\end{align}
	where we used triangle inequality in the second part. By taking expectation from both sides, we reach:
	\begin{align}\label{eq.12Edist2}
	&\mathds{E}_{\bm{g}}\mathrm{dist}^2(\bm{g},t\partial\|\cdot\|_{1,2,\bm{w}}(\bm{x}))=\nonumber\\
	&ks+\sum_{b\in \mathcal{B}}(tw_b)^2+\sum_{b\in {\overline{\mathcal{B}}}}\mathds{E}(\underbrace{\|\bm{g}_{\mathcal{V}_b}\|_2}_{\zeta}-tw_b)_{+}^2,
	\end{align}
	where $k=\tfrac{n}{q}$. and $\zeta^2:=\|\bm{g}_{\mathcal{V}_b}\|_2^2$ is distributed as chi-squared with $k$ degrees of freedom. Moreover,
	\begin{align}\label{eq.Ezetal12}
	&\mathds{E}(\zeta-tw_b)_{+}^2=2\int_{0}^{\infty}a\mathds{P}(\zeta^2\ge(a+tw_b)^2)da=\nonumber\\
	&\tfrac{2}{2^{\tfrac{k}{2}}\Gamma(\tfrac{k}{2})}\int_{0}^{\infty}\int_{(tw_b+a)^2}^{\infty}au^{\tfrac{k}{2}-1}e^{-\tfrac{u}{2}}du~da{\color{\change}=}\nonumber\\
	&\tfrac{2}{2^{\tfrac{k}{2}}\Gamma(\tfrac{k}{2})}\int_{(tw_b)^2}^{\infty}\int_{0}^{\sqrt{u}-tw_b}au^{\tfrac{k}{2}-1}e^{-\tfrac{u}{2}}da~du{\color{\change}=}\nonumber\\
	&\tfrac{1}{2^{\tfrac{k}{2}-1}\Gamma(\tfrac{k}{2})}\int_{tw_b}^{\infty}(u-tw_b)^2u^{k-1}e^{-\tfrac{u^2}{2}}du:=\tfrac{\phi(tw_b)}{2^{\tfrac{k}{2}-1}\Gamma(\tfrac{k}{2})},
	\end{align}
	where in the third line, the order of integration is changed and in the forth line, a change of variable is used. As a consequence, (\ref{eq.12Edist2}) becomes:
	\begin{align}\label{eq.12Edist2asli}
	&\mathds{E}_{\bm{g}}\mathrm{dist}^2(\bm{g},t\partial\|\cdot\|_{1,2,\bm{w}}(\bm{x}))=ks+\sum_{b\in \mathcal{B}}(tw_b)^2+\tfrac{\sum_{b\in \overline{\mathcal{B}}}\phi(tw_b)}{2^{\tfrac{k}{2}-1}\Gamma(\tfrac{k}{2})}.
	\end{align}
	By normalizing to the number of blocks $q$ and incorporating block prior information using $\bm{w}=\bm{D\omega} \in\mathbb{R}^q$ we reach
	\begin{align}
	&\mathds{E}_{\bm{g}}\mathrm{dist}^2(\bm{g},t\partial\|\cdot\|_{1,2,\bm{w}}(\bm{x}))=\nonumber\\
	&ks+\sum_{i=1}^{L}|\mathcal{P}_i\cap \mathcal{B}|t^2\omega_i^2+|\mathcal{P}_i\cap {\overline{\mathcal{B}}}|\tfrac{\phi(t\omega_i)}{2^{\tfrac{k}{2}-1}\Gamma(\tfrac{k}{2})}
	\nonumber\\
	&=q\bigg({\color{\change}k}\sigma+\sum_{i=1}^{L}\rho_i\big(\alpha_it^2\omega_i^2+\tfrac{(1-\alpha_i)\phi(t\omega_i)}{2^{\tfrac{k}{2}-1}\Gamma(\tfrac{k}{2})}\big)\bigg)\nonumber\\
	&=q\bigg(\sum_{i=1}^{L}\rho_i\big(\alpha_i(t^2\omega_i^2+k)+(1-\alpha_i)\tfrac{1}{2^{\tfrac{k}{2}-1}\Gamma(\tfrac{k}{2})}\phi(t\omega_i)\big)\bigg),\label{eq.Edist2part2}
	\end{align}
	where in the last line above, we benefited the fact that $\sigma=\sum_{i=1}^{L}\rho_i\alpha_i$.
\subsection{Proof of Proposition \ref{prop.uniqnessoptimal12weights}}\label{proof.uniqnessoptimal12weights}
	Define $\mathcal{C}:=\partial\|\cdot\|_{1,2}(\bm{x})$ and use Lemma \ref{lemma.mhat qsw} and \ref{lemma.Jb(nu)} to obtain:
	\begin{align}
	&\inf_{\bm{\omega}\in\mathbb{R}_{+}^L}\widehat{m}_{q,s,\bm{D\omega}}=\inf_{\bm{\omega}\in\mathbb{R}_{+}^L}\underset{{t\in\mathbb{R}_{+}}}{\inf}\Psi_{t,\bm{D\omega}}(\sigma,\bm{\rho},\bm{\alpha})=\inf_{\bm{\nu}\in\mathbb{R}_{+}^L}J_b(\bm{\nu}),\nonumber
	\end{align}
	where $\Psi_{t,\bm{D\omega}}(\sigma,\bm{\rho},\bm{\alpha})$ is defined in (\ref{eq.hatm-qsw}). Also, we used a change of variable $\bm{\nu}=t\bm{\omega}$ to convert multivariate optimization problem to a single variable optimization problem. Thus, the function $J_b(\bm{\nu})$ is obtained via the following equation:
	\begin{align}
	&J_b(\bm{\nu})=\sum_{i=1}^{L}\rho_i\big(\alpha_i(\nu(i)^2+1)+\tfrac{(1-\alpha_i)\phi(\nu(i))}{2^{\tfrac{k}{2}-1}\Gamma(\tfrac{k}{2})}\big).
	\end{align}
	By considering Lemma \ref{lemma.Jb(nu)} and $\bm{D}_t:=[\bm{1}_{\mathcal{V}_1},..., \bm{1}_{\mathcal{V}_q}]_{n\times q}[\bm{1}_{\mathcal{P}_1},..., \bm{1}_{\mathcal{P}_L}]_{q\times L}$, the function $J_b(\bm{\nu})$ is continuous and strictly convex and thus the unique minimizer can be obtained using $\nabla J_b(\bm{\nu})=\bm{0}\in\mathbb{R}^{L}$ which leads to
	\begin{align}
	2\alpha_i\nu^*(i)+\tfrac{2(1-\alpha_i)\phi'(\nu^*(i))}{2^{\tfrac{k}{2}-1}\Gamma(\tfrac{k}{2})}=0~:~i=1,..., L.
	\end{align}
\subsection{Proof of Proposition \ref{prop.error for mhat qsw}}\label{proof.mhatqsw}
	By the error bound (\ref{eq.errorbound}) and (\ref{eq.l12subdiff}), with $f(\bm{x})=\|\bm{x}\|_{1,2,\bm{w}}$ and $\bm{w}=\sum_{i=1}^{L}\omega_i\bm{1}_{\mathcal{P}_i} \in \mathbb{R}^q$, the numerator of (\ref{eq.errorbound}) is given by
	\begin{align}
	2\sup_{\bm{s}\in\partial\|\cdot\|_{1,2,\bm{w}}(\bm{x})}\|\bm{s}\|_2\le 2\sqrt{\sum_{i=1}^{q}w_b^2}=\nonumber\\
	2\sqrt{\sum_{i=1}^{L}|\mathcal{P}_i|\omega_i^2}=2\sqrt{\sum_{i=1}^{L}q\rho_i\omega_i^2}.\nonumber
	\end{align}
	Also, for the denominator we have:
	\begin{align}\label{eq.denomweight12}
	\tfrac{\|\bm{x}\|_{1,2,\bm{w}}}{{\|\bm{x}\|_2}}\le \sqrt{\sum_{i\in\mathcal{B}}w_i^2}=\sqrt{\sum_{i=1}^{L}|\mathcal{P}_i\cap \mathcal{B}|\omega_i^2}=\sqrt{\sum_{i=1}^{L}q\alpha_i\rho_i\omega_i^2},
	\end{align}
	where the first inequality in (\ref{eq.denomweight12}) follows from Cauchy--Schwartz inequality.
	The error bound (\ref{eq.errorbound}) for $\|\cdot\|_{1,2,\bm{w}}$ depends only on $\mathcal{D}(\|\cdot\|_{1,2,\bm{w}},\bm{x})$. Moreover, $\mathcal{D}(\|\cdot\|_{1,2,\bm{w}},\bm{x})$ only requires that $\|\bm{x}_{\mathcal{V}_b}\|_2=w_b~:~\forall b\in \mathcal{B}$. So, a vector
	\begin{align}
	\bm{z} = \left\{\begin{array}{lr}
	\|\bm{z}_{\mathcal{V}_b}\|_2=w_b, &  b\in \mathcal{B}\\
	0, & b\in \overline{\mathcal{B}}
	\end{array}\right\}\in \mathbb{R}^n\nonumber
	\end{align}
	can be chosen to satisfy equality in (\ref{eq.denomweight12}). Therefore, the error of obtaining the upper-bound of $m_{q,s,\bm{w}}$, i.e. $\widehat{m}_{q,s,\bm{w}}$ is
	\begin{align}\label{eq.rel2}
	\tfrac{2\sqrt{\sum_{i=1}^{L}q\rho_i\omega_i^2}}{q\sqrt{\sum_{i=1}^{L}q\alpha_i\rho_i\omega_i^2}}\le
	\tfrac{2}{q}\sqrt{\tfrac{1}{\underset{{i\in [q]}}\min\tfrac{|\mathcal{P}_i\cap \mathcal{B}|}{|\mathcal{P}_i|}}}
	\le \tfrac{2}{\sqrt{qL}},
	\end{align}
	in which the last inequality follows from the facts that $|\mathcal{P}_i\cap \mathcal{B}|\ge1$, $|\mathcal{P}_i|\le\tfrac{q}{L}$ for at least one $i\in[q]$ and thus $ \underset{{i\in [q]}}\min\tfrac{|\mathcal{P}_i\cap \mathcal{B}|}{|\mathcal{P}_i|}\ge \tfrac{L}{q}$. Further, the error of $\widehat{m}_{q,s,\bm{w}}$ from $m_{q,s,\bm{w}}$ is at most $\tfrac{2}{\sqrt{qL}}$. For the case of $\ell_{1,2}$, by using the facts $\omega_i=1 ~\forall i$ and $s=q\sum_{i=1}^L\rho_i\alpha_i$, it is straightforward to verify that $\tfrac{2\sqrt{\sum_{i=1}^{L}q\rho_i\omega_i^2}}{q\sqrt{\sum_{i=1}^{L}q\alpha_i\rho_i\omega_i^2}}=\tfrac{2}{\sqrt{q s}}$.
\subsection{Proof of Lemma \ref{lemma.Jb(nu)}}\label{proof.lem_convex}
\textit{Convexity}. Let $\bm{\nu}~,\tilde{\bm{\nu}}\in\mathbb{R}_{+}^L$ and $\theta\in[0,1]$ with $\bm{\upsilon}=\bm{D}_t\bm{\nu}$ and $\tilde{\bm{\upsilon}}=\bm{D}_t\tilde{\bm{\nu}}$. Then we have:
	\begin{align}\label{helpconvexity}
	&\forall \epsilon , \tilde{\epsilon}>0~\exists \bm{z} ,\tilde{\bm{z}}\in \mathcal{C} ~~\text{such that}\nonumber\\
	&\|\bm{g}-\bm{\upsilon}\odot \bm{z}\|_2\le \mathrm{dist}(\bm{g},\bm{\upsilon}\odot \mathcal{C})+\epsilon,\nonumber\\
	&\|\bm{g}-\tilde{\bm{\upsilon}}\odot \tilde{\bm{z}}\|_2\le \mathrm{dist}(\bm{g},\tilde{\bm{\upsilon}}\odot \mathcal{C})+\tilde{\epsilon}.
	\end{align}
	Since otherwise we have:
	\begin{align}
	&\forall \bm{z},\tilde{\bm{z}}\in \mathcal{C}:~\|\bm{g}-\bm{\upsilon}\odot \bm{z}\|_2>\mathrm{dist}(\bm{g},\bm{\upsilon}\odot \mathcal{C})+\epsilon,\nonumber\\
	&\|\bm{g}-\tilde{\bm{\upsilon}}\odot \tilde{\bm{z}}\|_2> \mathrm{dist}(\bm{g},\tilde{\bm{\upsilon}}\odot \mathcal{C})+\tilde{\epsilon}.
	\end{align}
	By taking the infimum over $\bm{z},\tilde{\bm{z}}\in \mathcal{C}$, we reach a contradiction. We proceed to show convexity of $\mathrm{dist}(\bm{g},(\bm{D}_t\bm{\nu})\odot \mathcal{C})$ by writing
	\begin{align}\label{eq.convexitydist}
	&\mathrm{dist}(\bm{g},(\theta\bm{\upsilon}+(1-\theta)\tilde{\bm{\upsilon}})\odot \mathcal{C})=\nonumber\\
	&\inf_{\bm{z} \in \mathcal{C}}\|\bm{g}-(\theta\bm{\upsilon}+(1-\theta)\tilde{\bm{\upsilon}})\odot \bm{z}\|_2\nonumber\\
	&\le\inf_{\bm{z}_1\in \mathcal{C},\bm{z}_2\in \mathcal{C}}\|\bm{g}-\theta\bm{\upsilon}\odot \bm{z}_1-(1-\theta)\bm{\upsilon}\odot \bm{z}_2\|_2\le\nonumber\\
	&\theta\|\bm{g}-\bm{\upsilon}\odot \bm{z}_1\|_2+(1-\theta)\|\bm{g}-\tilde{\bm{\upsilon}}\odot \bm{z}_2\|_2\le\nonumber\\
	&\theta \mathrm{dist}(\bm{g},\bm{\upsilon}\odot \mathcal{C})+(1-\theta)\mathrm{dist}(\bm{g},\tilde{\bm{\upsilon}}\odot \mathcal{C})+\epsilon+\tilde{\epsilon}.
	\end{align}
	Since this holds for any $\epsilon$ and $\tilde{\epsilon}$, $\mathrm{dist}(\bm{g},(\bm{D}_t\bm{\nu})\odot \mathcal{C})$ is a convex function. As the square of a non-negative convex function is convex, $J_{\bm{g}}(\bm{\nu})$ is a convex function. At last, the function $J(\bm{\nu})$ is the average of convex functions, hence is convex.
	In (\ref{eq.convexitydist}), the first inequality comes from the fact that $\forall \bm{z}_1,\bm{z}_2 \in \mathcal{C}~~\exists \bm{z}\in \mathcal{C}$:
	\begin{align}\label{eq.l12benefit}
	&\theta\bm{\upsilon}\odot \bm{z}_1+(1-\theta)\tilde{\bm{\upsilon}}\odot \bm{z}_2=\nonumber\\
	&\small{\left\{\bm{y}_{n\times1}:\begin{array}{lr}
		\bm{y}_{\mathcal{V}_b}=\big(\theta\bm{\upsilon}_{\mathcal{V}_b}+(1-\theta)\tilde{\bm{\upsilon}}_{\mathcal{V}_b}\big)\odot\tfrac{\bm{x}_{\mathcal{V}_b}}{\|\bm{x}_{\mathcal{V}_b}\|_2}, &  b\in \mathcal{B}\\
		\|\bm{y}_{\mathcal{V}_b}\|_2\le\theta\|\bm{\upsilon}\|_{\infty}\|{\bm{z}_1}_{\mathcal{V}_b}\|_2\nonumber\\
		+(1-\theta)\|\tilde{\bm{\upsilon}}\|_{\infty}\|{\bm{z}_2}_{\mathcal{V}_b}\|_2, & b\in {\overline{\mathcal{B}}}
		\end{array}\right\}}\nonumber\\
	&\in\small{\left\{\bm{y}_{n\times1}:\begin{array}{lr}
		\bm{y}_{\mathcal{V}_b}=\big(\theta\bm{\upsilon}_{\mathcal{V}_b}+(1-\theta)\tilde{\bm{\upsilon}}_{\mathcal{V}_b}\big)\odot\tfrac{\bm{x}_{\mathcal{V}_b}}{\|\bm{x}_{\mathcal{V}_b}\|_2}, &  b\in \mathcal{B}\\
		\|\bm{y}_{\mathcal{V}_b}\|_2\le\nonumber\\
		\big(\theta\|\bm{\upsilon}\|_{\infty}+(1-\theta)\|\tilde{\bm{\upsilon}}\|_{\infty}\big)\|{\bm{z}}_{\mathcal{V}_b}\|_2, & b\in \overline{\mathcal{B}}
		\end{array}\right\}}\nonumber\\
	&=(\theta \bm{\upsilon}+(1-\theta)\tilde{\bm{\upsilon}})\odot \bm{z}.
	\end{align}
	To verify (\ref{eq.l12benefit}), we argue by contradiction:
	\begin{align}
	&\forall \bm{z}\in \mathcal{C} ~\exists d\in \overline{\mathcal{B}}:~(\theta\|\bm{\upsilon}\|_{\infty}+(1-\theta)\|\tilde{\bm{\upsilon}}\|_{\infty}\big)\|{\bm{z}}_{\mathcal{V}_d}\|_2
	<\theta\|\bm{\upsilon}\|_{\infty}\nonumber\\
	&\|{\bm{z}_1}_{\mathcal{V}_d}\|_2+(1-\theta)\|\tilde{\bm{\upsilon}}\|_{\infty}\|{\bm{z}_2}_{\mathcal{V}_d}\|_2\le \theta\|\bm{\upsilon}\|_{\infty}+(1-\theta)\|\tilde{\bm{\upsilon}}\|_{\infty}.
	\end{align}
	Then, by taking $\bm{z}_{\mathcal{V}_d}=\bm{e}_i\in\mathbb{R}^{k}$ for some $i\in[k]$, we reach a contradiction.
	In the second inequality in (\ref{eq.convexitydist}), we used triangle inequality of norms. The third inequality uses (\ref{helpconvexity}).\par
	\textit{Strict convexity}. We show strict convexity by contradiction. If $J(\bm{\nu})$ was not strictly convex, there would be vectors $\bm{\nu},\tilde{\bm{\nu}}\in\mathbb{R}_{+}^L$ with $\bm{\upsilon}=\bm{D}_t\bm{\nu}, \tilde{\bm{\upsilon}}=\bm{D}_t\tilde{\bm{\nu}}$ and $\theta\in (0,1)$ such that
	\begin{align}\label{eq.strictconvex}
	\mathds{E}[J_{\bm{g}}(\theta\bm{\nu}+(1-\theta)\tilde{\bm{\nu}})]=\mathds{E}[\theta J_{\bm{g}}(\bm{\nu})+(1-\theta)J_{\bm{g}}(\tilde{\bm{\nu}})].
	\end{align}
	For each $\bm{g}$ in (\ref{eq.strictconvex}) the left-hand side is smaller than or equal to the right-hand side. Therefore, in (\ref{eq.strictconvex}), $J_{\bm{g}}(\theta\bm{\nu}+(1-\theta)\tilde{\bm{\nu}})$ and $\theta J_{\bm{g}}(\bm{\nu})+(1-\theta)J_{\bm{g}}(\tilde{\bm{\nu}})$ are almost surely equal (except at a set of measure zero) with respect to Gaussian measure. Moreover, we have
	\begin{align}\label{eq.J012}
	&J_{\bm{0}}(\theta\bm{\nu}+(1-\theta)\tilde{\bm{\nu}})=\mathrm{dist}^2(\bm{0},\big(\theta\bm{\upsilon}+(1-\theta)\tilde{\bm{\upsilon}}\big)\odot \mathcal{C})\le\nonumber\\
	&\inf_{\bm{z}_1,\bm{z}_2\in \mathcal{C}}\|\theta \bm{\upsilon}\odot \bm{z}_1+(1-\theta)\tilde{\bm{\upsilon}}\odot \bm{z}_2\|_2^2
	<\theta\inf_{\bm{z}_1\in \mathcal{C}}\|\bm{\upsilon}\odot \bm{z}_1\|_2^2+\nonumber\\
	&(1-\theta)\inf_{\bm{z}_2\in \mathcal{C}}\|\tilde{\bm{\upsilon}}\odot \bm{z}_2\|_2^2=\theta J_{\bm{0}}(\bm{\nu})+(1-\theta)J_{\bm{0}}(\tilde{\bm{\nu}}),
	\end{align}
	where the first inequality comes from (\ref{eq.l12benefit}) and the second inequality stems from the strict convexity of $\|\cdot\|_2^2$. From (\ref{eq.l12benefit}), it is easy to verify that the set $\bm{\nu}\odot \mathcal{C}$ is a convex set. The distance to a convex set e.g. $\mathcal{E}$ i.e. $\mathrm{dist}(\bm{g},\mathcal{E})$ is a $1$-lipschitz function (i.e. $|\mathrm{dist}(\bm{g},\mathcal{E})-\mathrm{dist}(\tilde{\bm{g}},\mathcal{E})|\le\|\bm{g}-\tilde{\bm{g}}\|_2~:~\forall~\bm{g},\tilde{\bm{g}}\in\mathbb{R}^n$) and hence continuous with respect to $\bm{g}$. Therefore, $J_{\bm{g}}(\bm{\nu})$ is continuous with respect to $\bm{g}$. So there exist an open ball around $\bm{g}=\bm{0}\in\mathbb{R}^n$ that similar to (\ref{eq.J012}), we may write the following relation for some $\epsilon>0$.
	\begin{align}
	&\exists \bm{u}\in\mathds{B}_{\epsilon}^n:~ J_{\bm{u}}(\theta\bm{\nu}+(1-\theta)\tilde{\bm{\nu}})<\theta J_{\bm{u}}(\bm{\nu})+(1-\theta)J_{\bm{u}}(\tilde{\bm{\nu}}).
	\end{align}
	The above statement contradicts with (\ref{eq.strictconvex}) and hence we have strict convexity. Continuity along with convexity of $J$ implies that $J$ is convex on the whole domain $\bm{\nu}\in\mathbb{R}_{+}^L$.
	
	\textit{Differentiability and continuity}. The function $J_{\bm{g}}(\bm{\nu})$ is continuously differentiable and the gradient for $\bm{\nu}\in\mathbb{R}_{++}^L$ is
	\begin{align}\label{eq.diffJg}
	&\nabla_{\bm{\nu}}J_{\bm{g}}(\bm{\nu})=\nonumber\\
	&\frac{\partial J_{\bm{g}}(\bm{\nu})}{\partial\bm{\nu}}=-2\bm{D}_t^T(\bm{D}_t\bm{\nu})^{\odot(-1)}\odot\mathcal{P}_{(\bm{D}_t\bm{\nu})\odot\mathcal{C}}(\bm{g})\nonumber\\
	&\odot(\bm{g}-\mathcal{P}_{\bm{D}_t\bm{\nu}\odot\mathcal{C}}(\bm{g})).
	\end{align}
	Continuity of $\frac{\partial J_{\bm{g}}(\bm{\nu})}{\partial\bm{\nu}}$ at $\bm{\nu}\in\mathbb{R}_{+}^L$ stems from the fact that the projection onto a convex set is continuous. For each compact set $\mathcal{I}\subseteq\mathbb{R}_{+}^L$ we have:
	\begin{align}
	&\mathds{E}\sup_{\bm{\nu}\in\mathcal{I}}\|\nabla_{\bm{\nu}} J_{\bm{g}}(\bm{\nu})\|_2\le\nonumber\\
	&2\|\bm{D}_t\|_{2\rightarrow2}\sqrt{q}(\sqrt{n}+2\sqrt{q}\big(\sup_{\bm{\nu}\in\mathcal{I}}\nu_{\max}\big))<\infty,
	\end{align}
	where $\nu_{\max}:=\underset{i\in[L]}{\max}~\nu(i)$. Therefore, we have
	\begin{align}
	\nabla_{\bm{\nu}} J(\bm{\nu})=\big(\frac{\partial}{\partial\bm{\nu}}\big)\mathds{E} J_{\bm{g}}(\bm{\nu})=\mathds{E}[\nabla_{\bm{\nu}} J_{\bm{g}}(\bm{\nu})] ~:~\forall \bm{\nu}\in\mathbb{R}_{+}^L,
	\end{align}
	where in the last equality, we used the Lebesgue's dominated convergence theorem. Also, continuity of $J(\bm{\nu})$ can be concluded from the continuity of its gradient.
	
	\textit{Attainment of the minimum}. Suppose that $\bm{\nu}\ge{\|\bm{g}\|_2}\bm{1}_{L\times 1}$. With this assumption, we may write:
	\begin{align}\label{eq.attain}
	&\mathrm{dist}(\bm{g},(\bm{D}_t\bm{\nu})\odot \mathcal{C})=\inf_{\bm{z}\in \mathcal{C}}\|\bm{g}-\bm{\upsilon}\odot \bm{z}\|_2\ge\nonumber\\
	&\inf_{\bm{z}\in \mathcal{C}}(\|\bm{\upsilon}\odot \bm{z}\|_2-\|\bm{g}\|_2)\ge\nu_{\min}-\|\bm{g}\|_2\ge0,
	\end{align}
	where in (\ref{eq.attain}), $\nu_{\min}:=\underset{i\in[L]}{\min}~\nu(i)$. By squaring (\ref{eq.attain}), we reach
	\begin{align}\label{eq.attainJ_g}
	J_{\bm{g}}(\bm{\nu})\ge (\nu_{\min}-\|\bm{g}\|_2)^2~~:~\forall \nu>{\|\bm{g}\|_2}\bm{1}_{L\times 1}
	\end{align}
	Using $\mathds{E}\|\bm{g}\|_2\ge\frac{n}{\sqrt{n+1}}$ \cite[Proposition 8.1]{foucart2013mathematical} and Marcov's inequality we obtain:
	\begin{align}
	\mathds{P}(\|\bm{g}\|_2\le\sqrt{n})\ge1-\sqrt{\frac{n}{n+1}}\nonumber.
	\end{align}
	Then we get:
	\begin{align}\label{eq.attainJ}
	&J(\bm{\nu})\ge\mathds{E}[J_{\bm{g}}(\bm{\nu})|\|\bm{g}\|_2\le\sqrt{n}]\mathds{P}(\|\bm{g}\|_2\le\sqrt{n})\nonumber\\
	&\ge(1-\sqrt{\frac{n}{n+1}})\mathds{E}\big[(\nu_{\min}-\|\bm{g}\|_2)^2|\|\bm{g}\|_2\le\sqrt{n}\big]\nonumber\\
	&\ge(1-\sqrt{\frac{n}{n+1}})(\nu_{\min}-\sqrt{n})^2,
	\end{align}
	where the first inequality stems from total probability theorem, the second inequality follows from (\ref{eq.attainJ_g}). From (\ref{eq.attainJ}), we find out that $J(\bm{\nu})>J(\bm{0})$ when $\nu>(2^{\frac{1}{4}}+1){\sqrt{n}}\bm{1}_{L\times1}$. Therefore, the unique minimizer of the function $J$ must occur in the interval $[\bm{0}, (2^{\frac{1}{4}}+1){\sqrt{n}}\bm{1}_{L\times1}]$.
\subsection{Proof of Theorem \ref{thm.l12weighted}}\label{proof.thm2}
	Using optimal weights, the upper-bound for the normalized number of measurements required for $\mathsf{P}_{1,2,\bm{D\omega}^*}^{0}$ to succeed is:
	\begin{align}
	&\widehat{m}_{q,s,\bm{w}^*}=\inf_{\omega\in\mathbb{R}_{+}^L}\widehat{m}_{q,s,\bm{D\omega}}=\nonumber\\
	&\sum_{i=1}^{L}[\inf_{\nu_i\in\mathbb{R}_{+}}\underbrace{\bigg(\tfrac{\|\bm{x}_{\mathcal{P}_i}\|_{0,2}}{q}(\nu_i^2+k)+(1-\tfrac{\|\bm{x}_{\mathcal{P}_i}\|_{0,2}}{q})\phi_B(\nu_i)\bigg)}_{\Psi_{\nu_i,\|\bm{x}_{\mathcal{P}_i}\|_{0,2}}(\tfrac{\|\bm{x}_{\mathcal{P}_i}\|_{0,2}}{q})}]\nonumber\\
	&=\sum_{i=1}^{L}\widehat{m}_{q,\|\bm{x}_{\mathcal{P}_i}\|_{0,2}}.
	\end{align}
	The expression in the bracket is the upper-bound for normalized number of measurements required for successful recovery of $\bm{x}_{\mathcal{P}_i}\in\mathbb{R}^n$ using $\mathsf{P}_{1,2}^{0}$ i.e. $\widehat{m}_{q,\|\bm{x}_{\mathcal{P}_i}\|_{0,2}}$. Thus, regarding the error bounds obtained in Proposition \ref{prop.error for mhat qsw}, the relation between $m_{q,s,\bm{D\omega}^*}$ and $m_{q,\|\bm{x}_{\mathcal{P}_i}\|_{0,2}}$ is given by \eqref{eq.rel1}.
\subsection{Proof of Proposition \ref{prop.robustness}}\label{proof.robust}
We have that $\lim\limits_{\alpha\rightarrow \alpha'}\tfrac{|\omega-\omega'|}{|\alpha-\alpha'|}=|\tfrac{\partial \omega}{\partial \alpha}|.$ By differentiating (\ref{eq.12optimalweights}), we reach
	\begin{align}
	\tfrac{\partial \omega}{\partial \alpha}=-\tfrac{\omega+\tfrac{1}{2^{k/2-1}\Gamma(k/2)}\int_{\omega}^\infty(u-\omega)u^{k-1}e^{-\tfrac{u^2}{2}}du}{\alpha+\tfrac{1-\alpha}{2^{k/2-1}\Gamma(k/2)}\int_{\omega}^\infty u^{k-1}e^{-\tfrac{u^2}{2}}du}.
	\end{align} 	
	Using (\ref{eq.12optimalweights}), the above equation reduces to:
	\begin{align}\label{eq.differen}
	\tfrac{\partial \omega}{\partial \alpha}=-\tfrac{\omega^22^{k/2-1}\Gamma(k/2)}{(1-\alpha)^2\int_{\omega}^{\infty}u^ke^{-\tfrac{u^2}{2}}du}.
	\end{align}
	By obtaining $\alpha$ from (\ref{eq.12optimalweights}) and replacing in (\ref{eq.differen}), we reach:
	\begin{align}
	&\tfrac{\partial \omega}{\partial \alpha}=-\tfrac{\Big(\int_{\omega}^\infty u^k e^{-\tfrac{u^2}{2}}du-\omega\int_{\omega}^\infty u^{k-1}e^{-\tfrac{u^2}{2}} +2^{\tfrac{k}{2}-1}\Gamma(\tfrac{k}{2})\omega\Big)^2}{2^{\tfrac{k}{2}-1}\Gamma(\tfrac{k}{2})\int_{\omega}^\infty u^k e^{-\tfrac{u^2}{2}}du}:=f(\omega).
	\end{align}
	After some simplification, $f(\omega)$ reduces to
	\begin{align}\label{eq.fomega}
	f(\omega)=\tfrac{\Big(\sqrt{2}\omega\big(\Gamma(\tfrac{k}{2})-\gamma(\tfrac{k}{2},\tfrac{\omega^2}{2})\big)+2\gamma(\tfrac{k}{2},\tfrac{\omega^2}{2})\Big)^2}{2\sqrt{2}\Gamma(\tfrac{k}{2})\gamma(\tfrac{k}{2},\tfrac{\omega^2}{2})}.
	\end{align}
	$f(\omega)$ implicitly depends on the accuracy $\alpha$ since $\omega$ is related to $\alpha$ by (\ref{eq.12optimalweights}). $f(\omega)$ can be further simplified to a function $c(k,\alpha)$ that only depends on $k$ and $\alpha$. This is accomplished by obtaining $\omega$ corresponding to any $\alpha\in [0,1]$ by (\ref{eq.12optimalweights}) and replacing the result into (\ref{eq.fomega}).
\ifCLASSOPTIONcaptionsoff
  \newpage
\fi

\bibliographystyle{ieeetr}
\bibliography{mypaperbibe1}

			\begin{IEEEbiography}
				[{\includegraphics[width=1in,height=1.25in,clip,keepaspectratio]{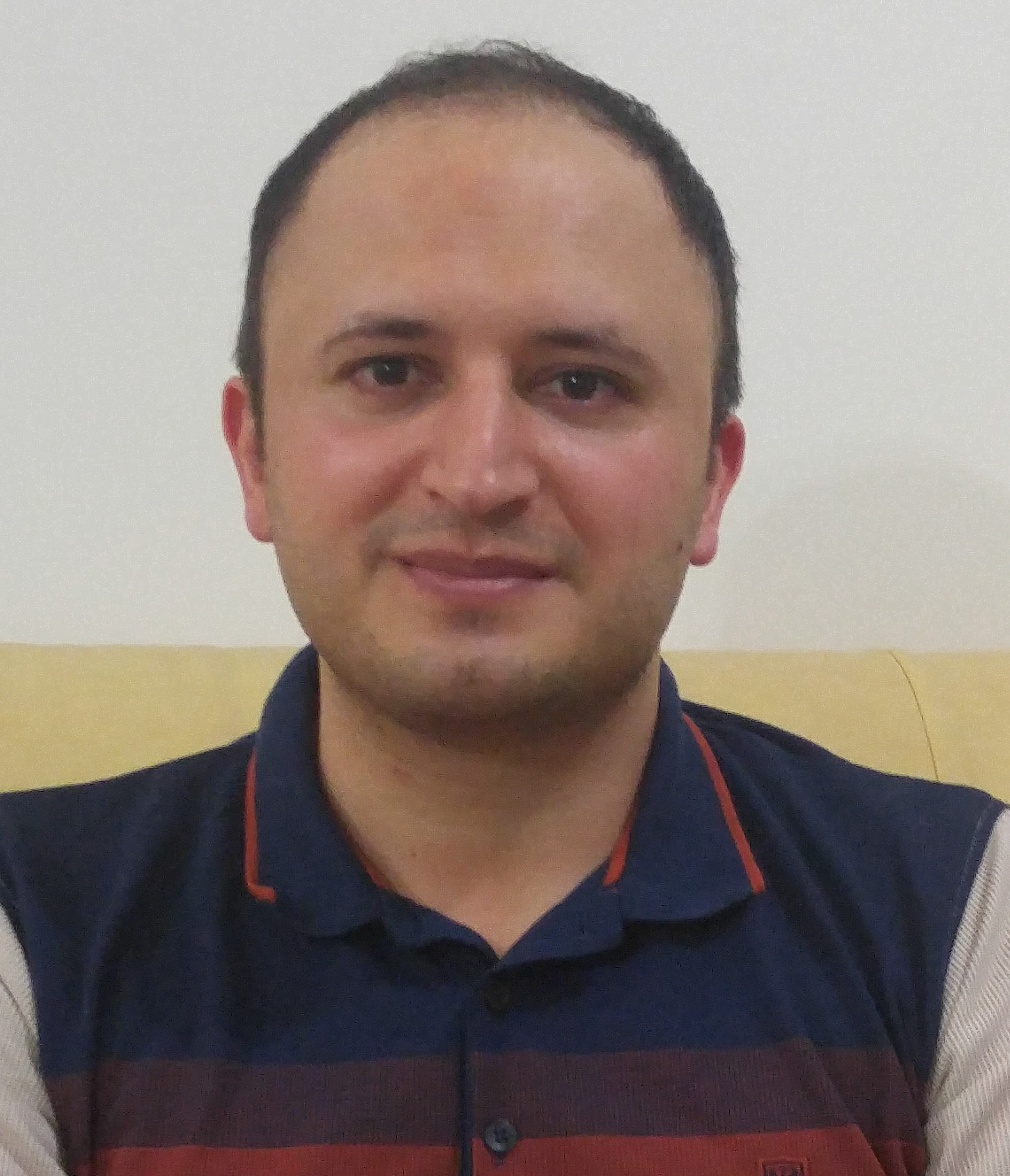}}]{\textbf{Sajad Daei}}
				received the B.Sc. degree in electronic engineering from Guilan University, Guilan, Iran, in 2011, and the M.Sc. degree in communication engineering from Sharif University of Technology, Tehran, Iran, in 2013. He is currently pursuing his Ph.D. at Iran University of Science \& Technology. His main research interests include convex optimization, compressed sensing and super resolution.
			\end{IEEEbiography}
			\begin{IEEEbiography}[{\includegraphics[width=1in,height=1.25in,clip,keepaspectratio]{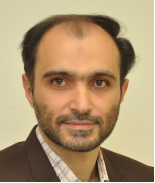}}]{\textbf{Farzan Haddadi}}was born in 1979. He received his B.Sc., M.Sc., and Ph.D. degrees in communication systems in 2001, 2003, and 2010, respectively, from Sharif University of Technology, Tehran, Iran. He joined Iran University of Science \& Technology faculty in 2011. His main research interests are array signal processing, statistical signal processing, subspace tracking, and compressed sensing.
			\end{IEEEbiography}

			\begin{IEEEbiography}[{\includegraphics[width=1in,height=1.25in,clip,keepaspectratio]{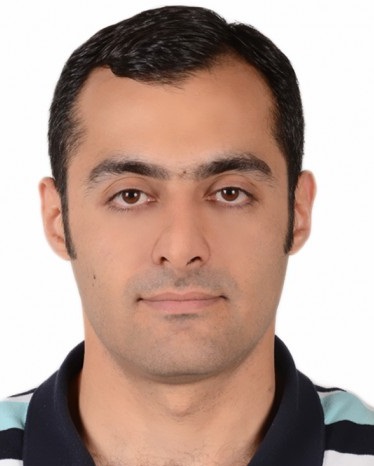}}]{\textbf{Arash Amini}}
received the B.Sc., M.Sc., and Ph.D. degrees in electrical engineering (communications and signal processing) and the B.Sc. degree in petroleum engineering (reservoir) from the Sharif University of Technology, Tehran, Iran, in 2005, 2007, 2011, and 2005, respectively. He was a Researcher with the \'Ecole Polytechnique f\'ed\'erale de Lausanne, Lausanne, Switzerland, from 2011 to 2013, working on statistical approaches toward modeling sparsity in continuous-domain. He joined Sharif University of Technology as an assistant professor in 2013, where he is now an associate professor since 2018. He has served as an associate editor of IEEE Signal Processing Letters from 2014 to 2018. His research interests include various topics in statistical signal processing, specially compressed sensing.			
			\end{IEEEbiography}

\end{document}